\documentclass[a4paper,11pt]{article}
\pdfoutput=1 

\usepackage{jheppub} 

\usepackage[T1]{fontenc} 
 
\usepackage{slashed}
\usepackage{booktabs}
\usepackage{feynmp}
\usepackage{multirow}
\usepackage{graphicx}
\usepackage[section]{placeins}
\allowdisplaybreaks
\usepackage{longtable}
\usepackage{tabu}

\renewcommand{\imath}{\mathrm{i}}
\newcommand{\muu}[1]{\mu_u^{\text{eff,}#1}}
\newcommand{\mud}[1]{\mu_d^{\text{eff,}#1}}

\title{\boldmath Exploring the Higgs sector of the MRSSM with a light scalar }

\author[a,1]{Philip Diessner,\note{Corresponding author.}}
\author[b]{Jan Kalinowski,}
\author[a,b]{Wojciech Kotlarski}
\author[a]{and Dominik St\"ockinger}

\affiliation[a]{Institut f\"ur Kern- und Teilchenphysik, TU Dresden\\
 01069 Dresden, Germany}
\affiliation[b]{Faculty of Physics, University of Warsaw,\\ Pasteura 5, 02093 Warsaw, Poland}

\emailAdd{philip.diessner@mailbox.tu-dresden.de}
\emailAdd{jan.kalinowski@fuw.edu.pl}
\emailAdd{wojciech.kotlarski@fuw.edu.pl}
\emailAdd{dominik.stoeckinger@tu-dresden.de}

\abstract{
In a recent paper we showed that the Minimal R-symmetric Supersymmetric Standard Model 
(MRSSM) can accommodate the observed 125~GeV Higgs boson as the lightest scalar of the model in 
agreement with electroweak precision observables, in particular with the W boson mass and 
$T$ parameter. Here we explore a scenario with the light singlet (and bino-singlino) state in 
which the second-lightest scalar takes the role of the SM-like boson with mass close to 125 GeV. 
In such a case the second-lightest Higgs state gets pushed up via mixing already at tree-level and 
thereby reducing the required loop correction. Unlike in the NMSSM, the light singlet is 
necessarily connected with a light neutralino which naturally appears as a promising dark matter 
candidate. We show that dark matter and LHC searches place further bounds on this scenario 
and point out parameter regions, which are viable and of interest for LHC Run II and upcoming 
dark matter experiments. }

\begin{document} 
\maketitle
\flushbottom

\newpage
\section{Introduction}
After the discovery of a Standard Model (SM)-like Higgs boson at the
LHC~\cite{Aad:2012tfa,Chatrchyan:2012xdj,atlascms}, it remains an open question whether 
there are additional
scalar particles, possibly even with smaller mass. Supersymmetric
models always predict the existence of additional scalars, and the
possibility of light scalars has been explored both in the minimal
supersymmetric standard model (MSSM) 
\cite{Heinemeyer:2011aa,Benbrik:2012rm,Bechtle:2012jw} and in its extensions, such as
the NMSSM which contains a gauge singlet field \cite{Ellwanger:2009dp,Maniatis:2009re}.
In the NMSSM a light singlet-like scalar is motivated since it pushes
 the tree-level value of the SM-like Higgs boson mass up towards the
observed value, reducing the need for large loop
corrections \cite{
Kang:2012sy,Vasquez:2012hn,Belanger:2012tt,King:2012tr,Badziak:2013bda,
Christensen:2013dra,Barbieri:2013nka,Cao:2013gba,Ellwanger:2014dfa,
Huitu:2014lfa,King:2014xwa,Jeong:2014xaa,Guchait:2015owa}.
The interplay of the Higgs sector and dark matter results in the MSSM and NMSSM
was further discussed in \cite{Bottino:2011xv,Boehm:2013gst,Belanger:2013pna,
Arbey:2013aba,Cao:2013mqa,Beskidt:2014oea,Han:2014nba}.

In the present paper we consider the possibility of a light,
singlet-like scalar in the minimal R-symmetric supersymmetric model,
the MRSSM \cite{Kribs:2007ac}. We show that it is necessarily
connected with a very light neutralino LSP, and we explore the
phenomenology of the Higgs sector and the related chargino/neutralino
sector, taking into account collider and dark matter constraints.

The MRSSM is quite distinct from the
usual MSSM and NMSSM and is motivated in several ways. Its essential
feature is an exact, global U(1)
R-symmetry \cite{Fayet:1974pd,Salam:1974xa}, under which SM-fields and 
superpartners have different charges. This symmetry is stronger than
R-parity, and it forbids Majorana masses for gauginos, the MSSM
Higgsino mass parameter $\mu$, and
all left-right sfermion mixings. As a result, several of
the most important experimental constraints on supersymmetry are
alleviated: contributions to CP- and flavour-violating
observables are suppressed even in presence of flavour violation in
the sfermion sector \cite{Kribs:2007ac,Dudas:2013gga}, and the
production cross section for squarks is reduced, making squarks below
the TeV scale generically compatible with LHC data \cite{Kribs:2012gx}.

The most exciting implication of R-symmetry from the point of view of
scalars is that additional scalar fields are not ad hoc but are
enforced by an N=2 supersymmetric structure of the gauge/gaugino
sector: for each gauge
group factor, there are gauge fields, Dirac instead of Majorana gauginos,
and scalar fields in the adjoint representation. The MRSSM therefore
contains sgluons -- colour-octet scalars, a scalar SU(2) triplet, and
a scalar singlet. This scalar singlet behaves quite differently and is
more strongly connected to the other sectors of the theory than the
singlet of the NMSSM.

Previous phenomenological investigations of the MRSSM have not
focused on a light singlet, but it has been shown that models with R-symmetry and/or
Dirac gauginos contain promising dark matter
candidates \cite{Belanger:2009wf,Chun:2009zx,Buckley:2013sca}, and the collider
physics of the extra, non-MSSM-like states has been studied 
\cite{Fox:2002bu,Plehn:2008ae,Choi:2008ub,Choi:2010gc,Choi:2010an,
Benakli:2012cy,ATLAS:2012ds,TheATLAScollaboration:2013jha,Kotlarski:2013lja}. 

Recently in Refs.~\cite{Diessner:2014ksa,Diessner:2015yna} the
lightest (SM-like) Higgs boson mass has been computed at the one-loop
and leading two-loop level and the W-boson mass has been computed
taking into account tree-level corrections to $\rho$ from the triplet
vacuum expectation value and  full one-loop corrections to muon decay.
It was shown that the measured value of the Higgs boson mass at the
LHC can be accommodated in the MRSSM even for top squarks as light as 1 TeV. 
The outcome of these calculations was not  obvious since in the MRSSM
the  lightest Higgs boson tree-level mass is typically reduced
compared to the MSSM due to mixing with the additional scalars, and
the loop corrections cannot be enhanced by stop mixing. However,
the new fields and couplings can give rise to the necessary large loop
contributions to the Higgs mass without generating too large a
contribution to the W-boson mass. For a similar analysis
see Ref.~\cite{Bertuzzo:2014bwa}, where also a welcome
reduction of the level of fine-tuning was found.

The plan of the present paper is as follows. In section \ref{sec:higgs}
we remind the reader on the relevant model details and will describe how
demanding a light singlet scalar constrains the parameter space of the model.
Sections \ref{sect:LHC} and \ref{sec:dm} study how LHC and dark matter bounds, respectively, 
are affected by these constraints.  To highlight features of  different regions of parameters, which 
are also in agreement with
LHC and dark matter searches, we define the new Benchmark points
given in Tab.~\ref{tab:BMP}. They feature many states well below 1 TeV, including the top squark masses.  
 The combination of all different experimental constraints
on the parameter space of the MRSSM is summarized in section \ref{sec:results} which also contains 
our conclusions.

\begin{table}[t]
\begin{center}
\begin{tabular}{lrrr}
\toprule
&BMP4&BMP5&BMP6\\
\midrule
$\tan\beta$  & $40$ & $20$ & $6$\\
$B_\mu$      & $200^2$ & $200^2$ & $500^2$\\
$\lambda_d$, $\lambda_u$ & $0.01,-0.01$ & $0.0,-0.01$ & $0.0,0.0$\\
$\Lambda_d$, $\Lambda_u$ & $-1,-1.2$ & $-1,-1.15$ & $-1,-1.2$\\
\midrule
$M_B^D$ & $50$&$44$&$30$\\
$m_S^2$ & $30^2$&$40^2$&$80^2$\\
$m_{R_u}^2$, $m_{R_d}^2$&& $1000^2,700^2$& \\
$\mu_d$, $\mu_u$&$130,650$&$400,550$& $550,550$\\
$M_W^D$&$600$ &$500$&$400$\\
$M_O^D$&& $1500$&\\
$m_T^2$, $m_O^2$&&$3000^2,1000^2$&\\
\midrule
$m_{Q;1,2}^2$, $m_{Q;3}^2$&$1500^2,700^2$&$1300^2,700^2$&$1400^2,700^2$\\
$m_{D;1,2}^2$, $m_{D;3}^2$&$1500^2,1000^2$&$1300^2,1000^2$&$1400^2,1000^2$\\
$m_{U;1,2}^2$, $m_{U;3}^2$&$1500^2,700^2$&$1300^2,700^2$&$1400^2,700^2$\\
$m_{L;1,2}^2$, $m_{E;1,2}^2$&$800^2$, $800^2$&$1000^2$, $1000^2$&$500^2$, $350^2$\\
$m_{L;3,3}^2$, $m_{E;3,3}^2$ & $800^2$, $136^2$& $1000^2$, $1000^2$& $500^2$, $95^2$\\
\midrule
$m_{H_d}$ &$1217^2$&$211^2$&$1042^2$\\
$m_{H_u}$&$-(767^2)$&$-(207^2)$&$-(201)^2$\\
$v_S$&$-64.9$&$-42.5$&$-56.1$\\
$v_T$&$-1.08$&$-1.2$&$-1.1$\\
\bottomrule
\end{tabular}
\end{center}
\caption{Benchmark points for the scenario discussed here. Dimensionful parameters are given in
GeV or GeV${}^2$, as appropriate.}
\label{tab:BMP}
\end{table}

\section{Higgs constraints}
\label{sec:higgs}
\subsection{Scenario description}
The essential field content and parameters of the MRSSM can be read
off from the superpotential
\begin{align}
\nonumber W = & \mu_d\,\hat{R}_d \cdot \hat{H}_d\,+\mu_u\,\hat{R}_u\cdot\hat{H}_u\,+\Lambda_d\,\hat{R}_d\cdot \hat{T}\,\hat{H}_d\,+\Lambda_u\,\hat{R}_u\cdot\hat{T}\,\hat{H}_u\,\\ 
 & +\lambda_d\,\hat{S}\,\hat{R}_d\cdot\hat{H}_d\,+\lambda_u\,\hat{S}\,\hat{R}_u\cdot\hat{H}_u\,
 - Y_d \,\hat{d}\,\hat{q}\cdot\hat{H}_d\,- Y_e \,\hat{e}\,\hat{l}\cdot\hat{H}_d\, +Y_u\,\hat{u}\,\hat{q}\cdot\hat{H}_u\, .
\label{eq:superpot}
 \end{align} 
The MSSM-like fields are the Higgs doublet superfields $\hat{H}_{d,u}$
and the quark and lepton superfields $\hat{q}$, $\hat{u}$, $\hat{d}$,
$\hat{l}$, $\hat{e}$. The Yukawa couplings are the same as in the
MSSM. The new fields are the doublets $\hat{R}_{d,u}$, which contain
the Dirac mass partners of the higgsinos and the corresponding Dirac higgsino mass parameters are
denoted as $\mu_{d,u}$. The singlet, the SU(2)-triplet and the color-octet chiral 
superfields, $\hat{S}$, $\hat{T}$ and $\hat O$,  contain the Dirac mass
partners of the usual gauginos. The superpotential contains Yukawa-like
trilinear terms involving the new fields; the associated parameters
are denoted as $\lambda_{d,u}$ for the terms involving the singlet
and $\Lambda_{d,u}$ for the terms involving the triplet; terms involving the octet are 
not allowed by R-symmetry. 
Further important parameters of the MRSSM are the Dirac mass
parameters  $M_{B,W,O}^D$ for the U(1), SU(2)$_L$ and SU(3)$_c$  
gauginos,  respectively, 
the soft
scalar mass parameters $m^2_{S,T,O}$ for the singlet, triplet and octet states, the soft mass parameters 
$m^2_{H_d,H_u,R_d,R_u}$ for the Higgs and R-Higgs bosons,  and the standard 
$B_\mu$ parameter and sfermion mass parameters, while the trilinear
sfermion couplings are forbidden by R-symmetry. The explicit form of
the soft SUSY breaking potential is given in Ref.~\cite{Diessner:2014ksa}.

We begin the  discussion with the neutral Higgs sector of the
MRSSM. There are four complex (electrically and R-charge) neutral scalar
fields, which can mix. Their real and imaginary parts,  denoted as
$(\phi_d,\phi_u,\phi_S,\phi_T)$ and $(\sigma_d,\sigma_u,\sigma_S,\sigma_T)$, correspond to the two 
MSSM-like Higgs doublets $H_{d,u}$, and the $N=2$ scalar superpartners of
the hypercharge and SU(2)$_L$ gauge fields, $S$ and $T$. Since their
vacuum expectation values, denoted as $v_{d,u,S,T}$ respectively, are assumed real, the real and 
imaginary components do not mix, and the mass-square matrix breaks into two 4x4 sub-matrices.

In the pseudo-scalar sector the MSSM-like states $(\sigma_d,\sigma_u)$ do not mix with 
the singlet-triplet  states $(\sigma_S,\sigma_T)$ and the mass-square matrix breaks further 
into two 2x2 sub-matrices, see  Appendix \ref{sect:app}. Therefore the neutral Goldstone boson and
one of the pseudo-scalar Higgs bosons $A$ with $m_A^2=2B_\mu/\sin2\beta$  appear as in 
the MSSM.  On the other hand, in the  
$(\sigma_S,\sigma_T)$ sector the mixing is negligible for a heavy triplet mass $m_T$ and the lightest 
pseudo-scalar state is almost a pure singlet with its  mass given by the soft parameter $m_S$.   
Therefore, to a good approximation  it decouples and is of no interest for the following discussion.

In the scalar sector all fields mix and  the full 4x4 tree-level mass-square matrix ${\cal M}^{\phi}$ 
in the weak basis $(\phi_d,\phi_u,\phi_S,\phi_T)$  has been given explicitly 
in Ref.~\cite{Diessner:2014ksa}.  It is diagonalized by a unitary rotation 
$Z^H{\cal M}^{\phi}{Z^H}^\dagger$ to the mass-eigenstate basis $(H_1,H_2,H_3,H_4)$, 
ordered with increasing mass.  
For the qualitative discussion we consider the limit in which the triplet decouples due to a high
soft breaking mass term $m_T$ (a necessary condition to suppress  the tree-level triplet 
contribution to the W boson mass and $\rho$ parameter), and in which the pseudoscalar Higgs mass
$M_A$ and the value of $\tan\beta=v_u/v_d$ become large.
In this limit the SM-like Higgs boson is dominantly given by the
up-type field $\phi_u$, and to  highlight the main effect of the mixing  in 
the light-scalar scenario it is enough to recall  the most relevant part of the mass-square matrix, 
namely the 2x2 sub-matrix corresponding to the $(\phi_u,\phi_S)$ fields only, which reads 
\begin{align}
\mathcal{M}^{\phi}_{u,S}&= 
\begin{pmatrix} 
m_Z^2 +\Delta m^2_{rad}& v_u \left(\sqrt{2} \lambda_u \muu{-} +g_1 M_B^D\right) \\
v_u \left(\sqrt{2} \lambda_u\muu{-} +g_1 M_B^D\right) \; &
4(M_B^D)^2+m_S^2+\frac{\lambda_u^2 v_u^2}{2} \;\\ 
\end{pmatrix}
\;,
\label{eq:hu-s-matrix}
\end{align} 
where we have included the dominant radiative correction $\Delta m^2_{rad}$ to the diagonal element of the doublet field $\phi_u$, and we have used the abbreviations
\begin{align}
\mu_i^{\text{eff,}\pm}&
=\mu_i+\frac{\lambda_iv_S}{\sqrt2}
\pm\frac{\Lambda_iv_T}{2}
,
&i=u,d.
\end{align}

The parameters $M_{B}^D$, $\mu_{u}$ and $\lambda_{u}$ appear
in the scalar potential due to supersymmetry.%
\footnote{In Ref~\cite{Martin:2015eca} it was recently shown that Dirac masses could also be 
generated without contributions to the scalar sector. 
Here, we keep with the usual D-term generation of these mass terms for the MRSSM.
}
From this approximation to the mass-square matrix we can draw several immediate
consequences  on the four most important parameters in the
light-singlet scenario in which $m_{H_1}<m_{H_2}\approx 125$ GeV.\\[2mm]
a) Singlet soft mass $m_S^2$:

The soft breaking parameter $m_S^2$ appears in the diagonal element
for the singlet field $\phi_S$ in the mass
matrix \eqref{eq:hu-s-matrix}. To ensure 
that the lightest scalar state is a singlet-like a direct upper limit
can be put on this parameter: 
 \begin{equation}
m_S^2< m_Z^2+\Delta m^2_{rad}\;,
\label{approxmS}
\end{equation}
Therefore the numerical value of $m_S^2$ has to 
be limited to be below $\approx (120\text{ GeV})^2$ \\[2mm]
b) Dirac bino-singlino mass $M_B^D$:

Due to supersymmetry, the Dirac mass parameter $M_B^D$ appears also in
the singlet diagonal element, and moreover in the off-diagonal
elements as well. Hence,  a direct upper limit can be derived
 \begin{equation}
(M_B^D)^2< \frac{m_Z^2+\Delta m^2_{rad}}{4}\approx (60\text{ GeV})^2\;,
\label{approxMDB}
\end{equation} similar to the one on
$m_S^2$ (however stronger) to keep the singlet-diagonal element smaller
than the doublet one. 
Of course, the bino-singlino mass parameter appears also in the neutralino
mass matrix and is the dominant parameter that  determines the LSP mass in
the light singlet/bino-singlino scenario. Therefore constraints from dark matter and direct LHC
searches for neutralinos will have a strong impact on the Higgs
sector.\\[2mm]
c) Higgsino mass $\mu_u$:

The higgsino mass parameter $\mu_u$ appears in the off-diagonal
element of the mass matrix in Eq.~\eqref{eq:hu-s-matrix} in combination 
with $\lambda_u$. However $\mu_u$ also enters as a parameter in  the
chargino/neutralino sector, and it is strongly limited from below by
the direct chargino searches and dark matter constraints. 
We recall that $\mu_u$ has a strong influence on the loop
corrections $\Delta m^2_{rad}$ to the doublet component $\phi_u$, so it can affect the
mixing additionally through that diagonal element. \\[2mm]
d) Yukawa-like parameter $\lambda_u$:

The Yukawa-like parameter $\lambda_u$ multiplies the higgsino mass parameter $\mu_u$ in the
off-diagonal element of Eq.~\eqref{eq:hu-s-matrix}. The off-diagonal
element cannot be too large in order to avoid a negative determinant
of the mass matrix and thus tachyonic states. In the simple limit
$v_S,m_Z\ll\mu_u$, avoiding tachyons leads to the following bound on
$\lambda_u$:
\begin{equation}
\left| \lambda_u
+\frac{g_1 M_B^D v_u}{\sqrt{2} \mu_u v_u}\right|
<
\frac{m_Z\sqrt{4(M_B^D)^2+m_S^2}}{\sqrt{2} \mu_u v_u}
\label{approxlambdau}
\end{equation}
There can be a cancellation between $\lambda_u$ and the $M_B^D$-term,
but since $\mu_u$ is much larger than all other appearing mass
parameters, the value of $\lambda_u$ is necessarily very small.

In summary, these simple considerations lead to the following
promising parameter hierarchies for a light singlet scenario:
\begin{align}
m_S, M_B^D &\quad < \quad
m_Z \quad <\quad \mu_u \, ,
&
|\lambda_u|\quad \ll \quad 1 \, .
\label{eq:hierarchy}
\end{align}
\subsection{Quantitative analysis and comparison to experiment}

In our phenomenological investigations we will of course not use the
approximations made before but we will always use the most precise
available expressions including higher order corrections. This is done using the 
\textit{Mathematica}~\cite{Mathematica} package 
\texttt{SARAH}~\cite{SA1,SA2,SA3,SA4,SA5,Goodsell:2014bna} and \texttt{SPheno}~\cite{SP1,SP2}. 
This approach was cross-checked, as described in Refs.~\cite{Diessner:2014ksa,Diessner:2015yna}, 
using \texttt{FlexibleSUSY}\cite{FlexibleSUSY}, \texttt{FeynArts} and \texttt{FormCalc}~\cite{FA,Hahn:1998yk,FC}.

\begin{figure}
\includegraphics{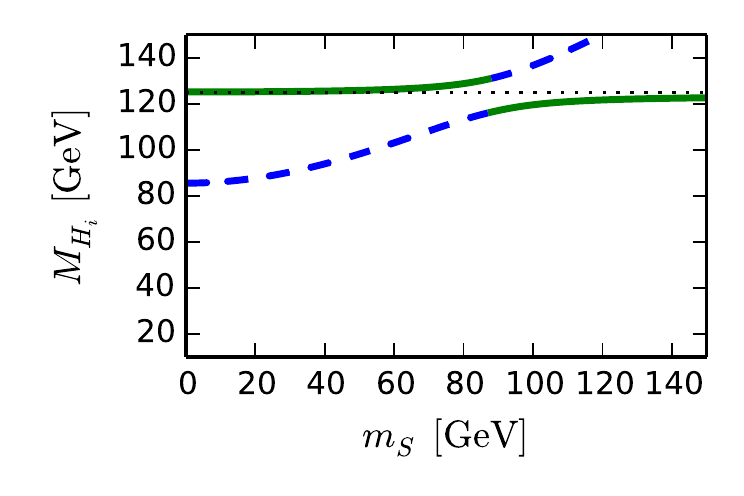}
\includegraphics{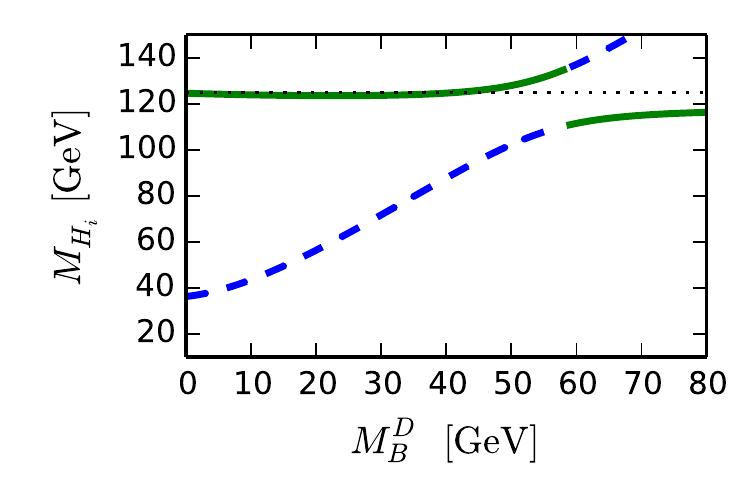}\\
\includegraphics{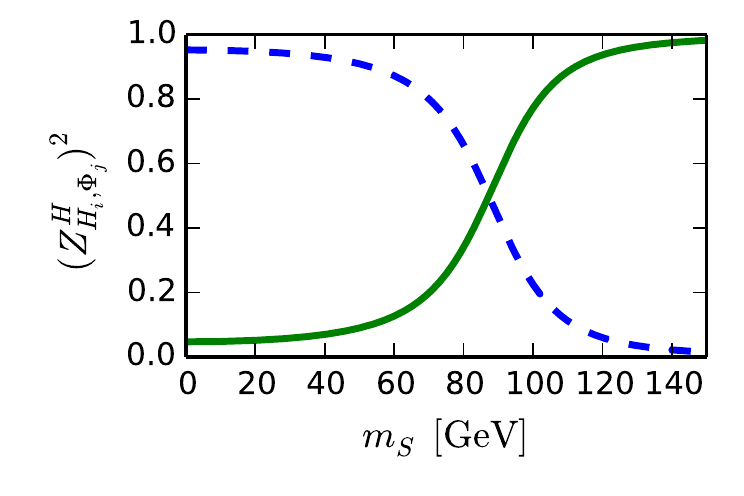}
\includegraphics{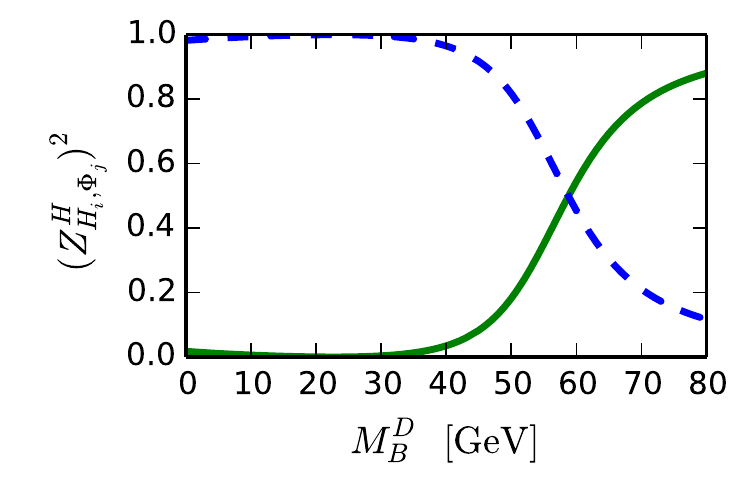}\\
\includegraphics{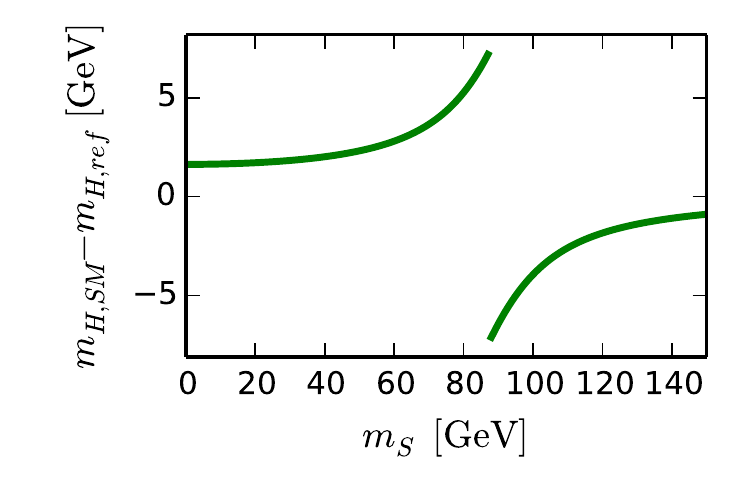}
\includegraphics{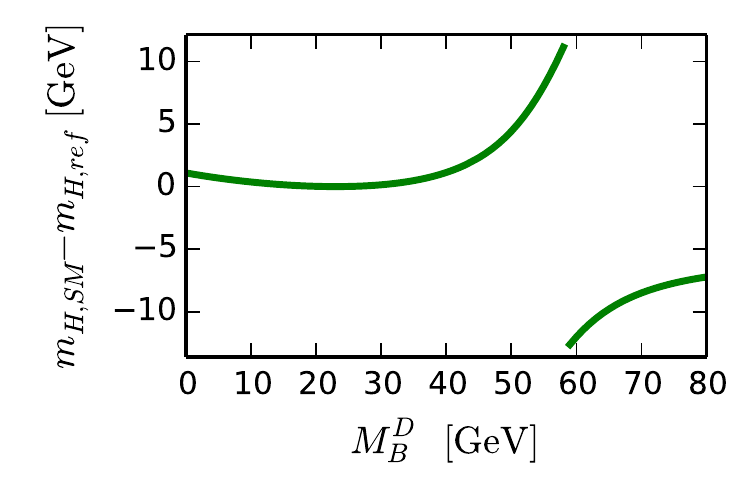}\\
\caption{Level crossing of lightest and second-lightest Higgs state for BMP5, depending on the dimensionful parameter $m_S$ ($M_B^D$) on the left (right). 
Top row shows the masses of the lightest and
second-lightest mass eigenstate, the green full line is more SM-like, while the blue dashed line is singlet-like.The second line 
shows the mixing parameter as described in the text. The blue dashed line is the singlet content of the lightest mass eigenstate, the green full line the doublet content.
The bottom row gives the relative mass shift compared to a reference mass, where the mixing between
singlet and doublet vanishes.}
\label{img:exclusion3}
\end{figure}
Fig.~\ref{img:exclusion3} shows a quantitative analysis of the mixing
of the two lightest Higgs states, as a function of the two relevant
parameters $m_S$ (left column) and $M_B^D$ (right column). The other
parameters are chosen as in  BMP5, given in Tab.~\ref{tab:BMP}. They satisfy
the hierarchy of Eq.~\eqref{eq:hierarchy}; the precise parameter choices
will be motivated in the subsequent sections. 

The first two lines in the figure show the masses of the two lightest
Higgs states (computed to two-loops as described in 
Ref.~\cite{Diessner:2015yna}) and the singlet component of these two
states, measured by the square of the corresponding mixing matrix
elements $(Z^H_{H_1,\phi_S})^2$ and $(Z^H_{H_2,\phi_S})^2$.
As expected from the discussion above, as long as the approximate
inequalities Eq.~\eqref{approxmS} and Eq.~\eqref{approxMDB} are satisfied, the
lightest state is significantly lighter than $m_Z$ and has a high
singlet component. Once $m_S$ or $M_B^D$ become heavier, the lightest
state becomes mainly a doublet-like (and hence SM-like) and its mass
approaches a limit, which is here below 120~GeV.
Which of the parameter regions is phenomenologically viable will be
determined below.

The third line of plots in Fig.~\ref{img:exclusion3} quantifies one of
the major motivations for the light singlet scenario --- the
enhancement of the SM-like Higgs boson mass due to the tree-level push resulting 
from mixing with the scalar state.  The plots show the
difference of the actual value of this mass and the value in a
reference case, in which all Higgs bosons except the SM-like one are
very heavy. 
In the light-singlet case, the upward shift can amount to more than
10~GeV. The upward shift is particularly strong for non-vanishing Dirac
mass $M_B^D$, since this parameter also appears in the off-diagonal
element of the mass matrix Eq.~\eqref{eq:hu-s-matrix}.

\begin{figure}[t]
\includegraphics{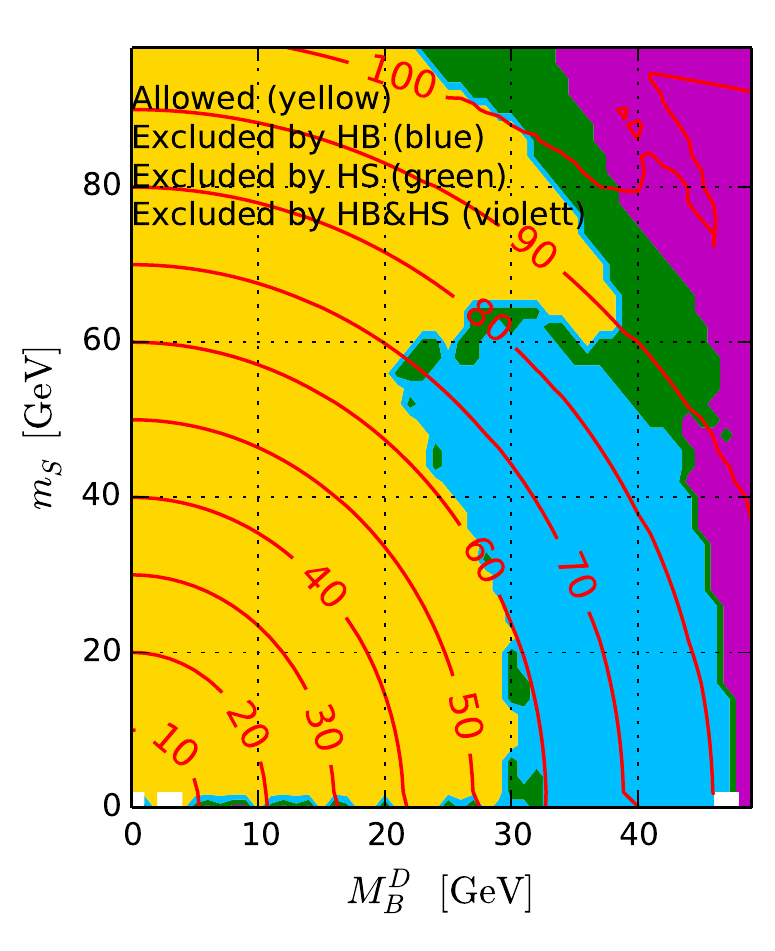}
\includegraphics{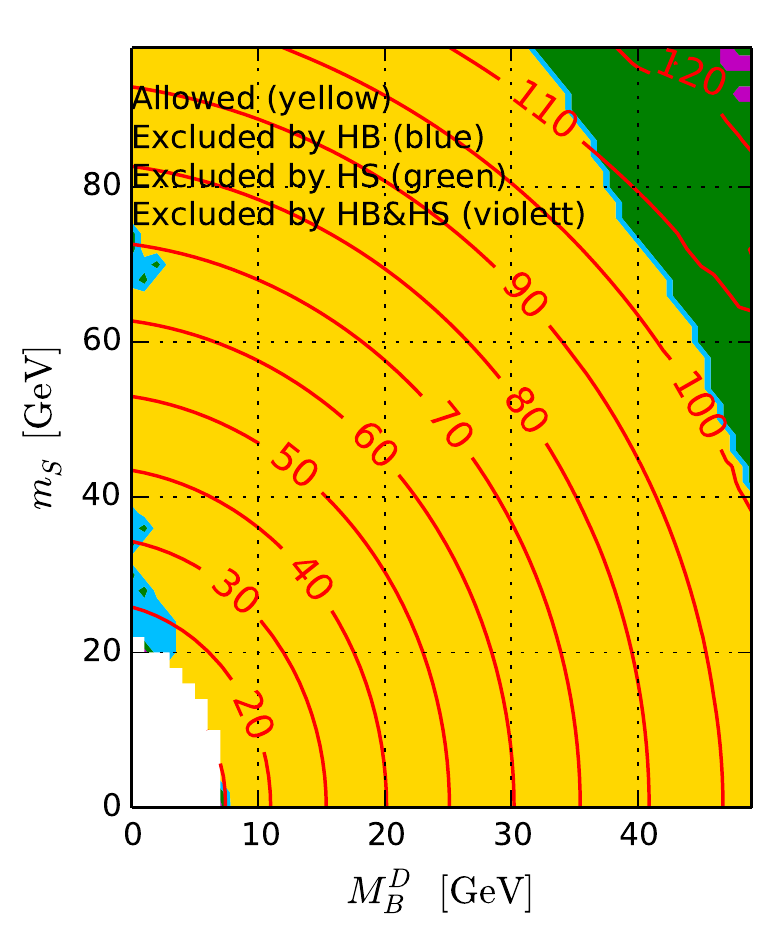}
\caption{Exclusion plot using \texttt{HiggsSignals-1.3} and \texttt{HiggsBounds-4.2};
Higgs masses are calculated at two loops.
Scanned over $\Lambda_u$ and projected onto the plane. If there is one $\Lambda_u$ for
a point giving non-exclusion, it is shown as non-excluded. Scanning from $-1.5<\Lambda_u<0$, while $\lambda_u=0/-0.01$ (left,right). 
Red contours show the lightest Higgs boson mass.}
\label{img:exclusion1}
\end{figure}

Now we compare the light-singlet scenario illustrated in
Fig.~\ref{img:exclusion3} with experimental data. It is clear that
there are two kinds of direct constraints on the two lightest scalars
and their mixing. On the one hand, there must be a SM-like state which
can be identified with the Higgs boson observed at LHC, with mass
compatible with the observed value and with sufficiently small singlet
mix-in in order to agree with the observed Higgs signal strengths and branching
ratios. On the other hand, the state that is lighter than the
one observed at the LHC must be sufficiently singlet-like in order to
evade limits from direct searches for light scalars, especially the LEP searches. 
Roughly speaking, for light singlet masses these constraints 
restrict the mixing to around $(Z^H_{H_1,\phi_u})^2<0.02$, except for
the region, where LEP saw an upward fluctuation. Here the mixing
is restrained to $(Z^H_{H_1,\phi_u})^2<0.15$.

Both types of constraints can be precisely analyzed with the help of
\texttt{HiggsBounds}-4.2 and \texttt{HiggsSignals}-1.4,
respectively \cite{HB1,HB2,HB3,HB4,HS1,HS2}. 
Here, we pass the effective couplings written by \texttt{SPheno}
to both tools for each parameter
point studied. 
With  \texttt{HiggsBounds} the option \textit{LandH} is used to
check the MRSSM Higgs sector against all experimental constraints included with the program.  
For \texttt{HiggsSignals} all peak
observables available are used (option \textit{latestresults} and \textit{peak}), 
the mass uncertainties are described using a Gaussian shape and we estimate the
uncertainty of the SM-like Higgs boson mass with 3~GeV. 

The interpretation of the \texttt{HiggsBounds} and \texttt{HiggsSignals}
results is done in a simplified way. A parameter point is excluded
by \texttt{HiggsBounds} by using the 95\% CL limit of its output 
and by \texttt{HiggsSignals} when its
approximately calculated p-value is smaller than 0.05.
A point is excluded by both when both constraints apply.
For a more complete statistical treatment of extended Higgs sectors containing 
light singlet scenarios
for the singlet-extended SM and NMSSM see 
Refs.~\cite{Robens:2015gla} and \cite{Domingo:2015eea}, respectively.
In the case of the MRSSM this approach is left for future work.

Fig.~\ref{img:exclusion1} shows the resulting excluded and allowed regions
in the plane of $m_S$ and $M_B^D$ for two different values of
$\lambda_u$. The parameter $\Lambda_u$ is adjusted to ensure the
correct mass of the observed Higgs boson; the remaining parameters are
fixed as in BMP5.
The combination of the two plots shows that
complete range of $m_S$ and $M_B^D$ discussed before with simple
approximations of the tree-level scalar mass matrix is viable,
except of the region of level crossing and very strong mixing. 
We find that the light-singlet scenario can be realized in the MRSSM
for $m_S<100$~GeV and $M_B^D<55$~GeV.
However, the two plots also show very high sensitivity to the small
$\lambda_u$, which appears essentially in combination with the very
small term $\propto M_B^D/\mu_u$, see \eqref{approxlambdau}. 
Changing it from zero to $(-0.01)$ leads to strong shifts of the
allowed regions.

\begin{table}[t]
\begin{center}
\begin{tabular}{lrrr}
\toprule
&BMP4&BMP5&BMP6\\
\midrule
$m_{H_1}$ &$100$&$94$&$95$\\
$m_{H_2}$&$125.8$&$125.5$&$125.8$\\
\texttt{HiggsSignals} p-value& $0.75$&$0.76$&$0.72$\\
Allowed by \texttt{HiggsBounds}&$\checkmark$&$\checkmark$&$\checkmark$\\
$m_W$ &$80.384$&$80.392$&$80.404$\\
\midrule
\bottomrule
\end{tabular}
\end{center}
\caption{Higgs sector observables: masses of the Higgs bosons, \texttt{HiggsSignals} p-value, 
\texttt{HiggsBounds} check and, for completeness, the  W boson mass for the benchmark points. 
Dimensionful parameters are given in GeV.}\label{tab:HiggsObs}
\end{table}

The main novel feature of the light-singlet scenario in the MRSSM is
the restriction on the Dirac bino-singlino mass $M_B^D$. This restriction has
no counterpart e.g.\ in the NMSSM.
Obviously this limit affects the neutralino sector.
Combined with LEP bounds on chargino and neutralino production it
leads to the conclusion that the LSP is a Dirac bino-singlino neutralino
with mass related to $M_B^D$ and thus limited from above.
This limit on the LSP mass will provide strong constraints from and
correlations to LHC processes and dark matter studies. 

With these restrictions on $m_S$ and $M_B^D$, the decay of the Z and SM-like Higgs boson into
the LSP and singlet-like Higgs boson, if kinematically allowed, needs to be sufficiently
suppressed so that theoretical prediction of invisible width of both particles is
in agreement with experiment. 
In the mass region of interest, $m_{H_1}<62$~GeV, \texttt{HiggsBounds} puts very stringent 
limits on the singlet-doublet mixing. As the decay of the SM-like Higgs boson into the light 
(pseudo-)scalar is related to this mixing, contributions of these decay channels 
are strongly reduced for realistic scenarios.
Decays of Z (Higgs) bosons to LSPs are induced by the couplings to the (R-)higgsino 
(higgsino+bino or R-higgsino+singlino) parts of the LSP. 
This is mainly mediated by gauge couplings and will be suppressed by the weak
(R-)higgsino--singlino-bino--mixing. We will show that it is also important
for direct detection limits from dark matter searches. These searches put such a strong
bound on the mixing so that the decay to the LSP will be automatically suppressed.

Table \ref{tab:HiggsObs} summarizes the Higgs sector observables for our benchmark points, 
as well as the predicted W boson mass.

\section{LHC constraints}
\label{sect:LHC}
As was discussed in the previous section, demanding the singlet state to give the lightest Higgs 
boson yields a direct upper limit on the Dirac bino mass $M_B^D$. 
A light bino-like neutralino, and more generally light weakly
interacting particles, are promising in two respects. They might lead
to observable signals at the LHC, and they might explain the observed
dark matter relic density. At the same time, the light-singlet
scenario is constrained by existing data from LHC and dark matter
searches. In the present section we study the constraints from LHC in
detail. We only consider constraints on weakly interacting particles,
i.e.\ on charginos, neutralinos, and sleptons. All
strongly interacting particles are not relevant for the electroweak
and dark matter phenomenology studied here and are assumed to be heavy
enough to evade limits. 

So far the negative results of LHC searches for new particles have
been interpreted within  
the MSSM, or a number of simplified models which include a pair of charginos, 
the neutralinos and gluinos are of the Majorana type,  and 
left- and right-chiral sfermions that can mix. In general, care has to be taken in
reinterpreting the LHC bounds in the context of the
MRSSM, since the MRSSM differs from the MSSM in the number of degrees
of freedom, the Dirac versus Majorana nature of neutralinos, the
mixing patterns, and since some processes are forbidden in the MRSSM
due to the conserved R-charge.

\subsection{Constraints from slepton searches}
We first discuss briefly the simplest case of constraints following from slepton searches. In
both the MSSM and the MRSSM, sleptons can be produced in pairs through
Drell-Yan processes. In the MRSSM, slepton left-right mixing vanishes,
which is not an essential difference to the MSSM, where the mixing is
typically assumed to be small. In both models, light sleptons will
decay directly to the corresponding leptons and the bino-like
neutralino LSP.  In such a case the Dirac or Majorana nature of the LSP does not matter and,   
hence, the MSSM exclusion bounds for light sleptons can be applied
directly to the MRSSM case.  Recently the ATLAS Collaboration derived the exclusion limits in 
the $(m_{\tilde{\ell}_{L,R}},m_{\tilde{\chi}^0_1})$ 
parameter space from the analyses of the selectron and smuon pair-production 
processes, see Figs.~8 (a) and (b) of Ref.~\cite{Aad:2014vma}.%
\footnote{Similar limits are also given by the CMS collaboration~\cite{Khachatryan:2014mma}.}
These exclusion limits can be specialized to our case, in which
Eq.~(\ref{approxMDB}) implies an upper limit of $\approx55$ GeV on the
LSP mass. We then find that left-handed slepton masses below
$\approx300$ GeV are excluded. For the right-handed slepton masses the
bound is somewhat weaker (around $250$ GeV) due to smaller production cross section. 
Further, a small corner
at a lower right-handed slepton mass $\sim 100$  GeV with the LSP masses in the
range  40-55~GeV is still allowed.
With the current experimental reach the direct production of staus with the cross
sections predicted by supersymmetry can not be excluded, see Ref.~\cite{Aad:2014yka}.

The remainder of the present section is devoted to the more
interesting case of neutralino and chargino
searches.

\subsection{Qualitative description of the MRSSM chargino/neutralino sector}

The chargino and neutralino sectors of the MSSM and the MRSSM are
quite different. In the MRSSM, the neutralinos are Dirac fermions
composed of eight Weyl spinors
$ {\xi}_i=({\tilde{B}}, \tilde{W}^0, \tilde{R}_d^0, \tilde{R}_u^0)$
and 
$\zeta_i=(\tilde{S}, \tilde{T}^0, \tilde{H}_d^0, \tilde{H}_u^0) $. The
four mass eigenvalues are dominantly given by the four independent
mass parameters $(M_B^D,M_W^D,\mu_d,\mu_u)$, see Appendix \ref{sect:app}. 
In contrast, in the MSSM
the neutralinos are Majorana fermions, and there is only a single
higgsino mass parameter $\mu$, so that two neutralino masses are
approximately degenerate.

The MRSSM comprises four different charginos, with masses determined
by the wino and the two higgsino mass parameters, see Appendix \ref{sect:app}.  In
contrast, the MSSM contains only two charginos.

Table \ref{img:lhc_mssm_mrssm}  gives an overview of possible
observed signatures of sets of charginos and neutralinos (listed in the first
column). It then shows the corresponding possible interpretations of
the signatures in the MSSM (second column) and in the MRSSM (third
column), which are very different as can be understood on the basis of the previous remarks.
The table is more general than
necessary for the purposes of the present paper, and it can serve as a
basis of further investigations of the LHC phenomenology of the
MRSSM.

In general, as the number of degrees of freedom for neutralinos and charginos is twice
as large in the MRSSM than in the MSSM one could  naively expect an
enhancement of the production cross section by a factor of
four. However, all MRSSM charginos and neutralinos carry non-vanishing
R-charge. Due to R-charge conservation, therefore,  only half of all
final state combinations is allowed. 
Furthermore, the new genuine MRSSM 
states ($\tilde{R}_d, \tilde{R}_u,\tilde{S}, \tilde{T}$) do not 
interact at tree level with  fermions, sfermions or gluons. Hence the
situation is more complicated and has to be analyzed channel by channel.

Further differences between MRSSM and MSSM exist in the
couplings of charginos and neutralinos and therefore in the decay
branching ratios. For example, in the MSSM the higgsino mass parameter $\mu$ 
induces a maximal mixing between the up- and down-higgsino so that both will mix almost
similarly  with the bino and wino, implying an appreciable decay rates
of both higgsinos to sleptons, if kinematically available.
This is different in the MRSSM. R-symmetry does not allow 
a parameter which induces  mixing between the up- and
down-(R-)higgsinos, instead separate mass parameters $\mu_d$ and $\mu_u$
are needed. Then, up- and down-(R-)higgsino states separately will be good
approximations to the corresponding mass eigenstates. 
For $\tan\beta \gg 1$ this also means that the
mixing between down-(R-)higgsino with the wino-triplino and bino-singlino states
will be strongly suppressed. 
As a result, for large $\tan\beta$ the down-(R-)higgsino has
appreciable couplings only to staus, which are not shared with
the up-(R-)higgsino.

It is also important to consider decays of charginos and neutralinos
into the SM-like Higgs boson $H_2$ observed at the LHC or into the light singlet $H_1$,
present in our scenario. These decays are influenced by the
$\lambda/\Lambda$ parameters. Large $\Lambda$ are needed to generate large enough loop contributions 
to the mass of the SM-like Higgs boson  and $\lambda$ needs to be  small in the 
light singlet scenario studied here.
Therefore, the decay of the wino-triplino  to the SM-like Higgs
boson and LSP, if kinematically allowed, will be enhanced by $\Lambda$,
while contributions from $\lambda$ to the higgsinos decaying to the SM-like Higgs boson and LSP
are small. 

When the chargino/neutralino decay to the SM-like Higgs boson is kinematically not possible, 
decays to the
light singlet in our scenario still might open. 
Therefore, it can be a competing channel to the
decay to the Z boson above the Z boson mass or the dominant decay channel below it.
For higher neutralino masses the branching ratio to the light singlet will
be sub-dominant to the one to the SM-like Higgs boson, but might still be non-negligible, depending on the mixing between singlet
and doublets.

\begin{table}[t!]
\begin{tabular}{lcc}
\toprule
experimental signature & {\bf MSSM} & {\bf MRSSM}\\
\midrule
1 charged and  & \multirow{2}{4cm}{wino-like states} &\multirow{2}{6cm}{either up- or down-(R-)higgsino-like states } \\
1 neutral of similar mass & & \\
\midrule
2 charged and  & \multirow{2}{4cm}{\centering-} &\multirow{2}{6cm}{wino-triplino-like states} \\
1 neutral of similar mass & & \\
\midrule
1 charged and  & \multirow{2}{4cm}{higgsino-like states} &\multirow{2}{6cm}{\centering- } \\
2 neutral of similar mass & & \\
\midrule
2 charged and  & \multirow{2}{4cm}{\centering-} &\multirow{2}{6cm}{either up- or down-(R-)higgsino-like states } \\
2 neutral of similar mass & & \\
\midrule
2 charged and  & \multirow{2}{4cm}{all states} &\multirow{2}{6cm}{\centering- } \\
3 neutral of similar mass & & \\
\midrule
3 charged and  & \multirow{2}{4cm}{\centering-} &\multirow{2}{6cm}{wino-triplino-like and either up- or down-(R-)higgsino-like states} \\
2 neutral of similar mass & & \\
\midrule
4 charged and  & \multirow{2}{4cm}{\centering-} &\multirow{2}{6cm}{all states} \\
3 neutral of similar mass & & \\
\bottomrule
\end{tabular}
\caption{Possible discovery scenarios at colliders of different sets of particles from the neutralino-chargino sector 
and the corresponding dominant gauge eigenstates for the MSSM and MRSSM. This is assuming the light-singlet-scenario in the MRSSM, which leads
to a light bino-singlino as LSP candidate.}
\label{tab:lhc-scenarios}
\end{table}

\subsection{Setup for recasting LHC limits for the MRSSM}
After having given an overview on the qualitative details of 
electro-weak production in the MRSSM,
we now turn to a more in-depth analysis, taking the experimental input into account
in a systematic way. 

In recent years different computational tools emerged, which try to automatize
the study of beyond the Standard Model physics at the LHC in a rather generic way using
standardized interfaces.
This allows us to take  the model differences between the MRSSM and
the MSSM/simplified models studied by the experiments into account and directly 
calculate bounds from the LHC on
the MRSSM parameter space. In the following, we give a
description of the set of tools used to for this.

Using the \texttt{UFO}~\cite{Degrande:2011ua} output produced by \texttt{SARAH} with
\texttt{Herwig++}-2.7~\cite{Bahr:2008pv,Bellm:2013hwb}%
\footnote{\texttt{Herwig++}-2.7 is used with default settings, underlying event tune 
(UE-EE-5-MRST~\cite{Seymour:2013qka}) and PDF set ($LO^*$ from \cite{Sherstnev:2007nd}).
} we simulate the LO production of electro-weak
sparticles at the LHC and compare it with 8 TeV data with \texttt{CheckMATE}-1.2.0~\cite{Drees:2013wra}.%
\footnote{With \texttt{CheckMATE} we make use of \texttt{Delphes}~3~\cite{deFavereau:2013fsa}, 
\texttt{FastJet}~\cite{Cacciari:2011ma}, 
the anti-$k_t$ clustering~\cite{Cacciari:2005hq,Cacciari:2008gp},
and the $CL_s$ description~\cite{Read:2002hq}.}
\texttt{CheckMATE} includes ATLAS analyses, which can be sensitive 
to the final state of these processes, when several leptons appear as
decay products. Specifically, we take into account  the 
analyses implemented in \texttt{CheckMATE} for two~\cite{TheATLAScollaboration:2013hha,Aad:2014vma}, 
three~\cite{ATLAS:2013rla,Aad:2014nua} and four and
more leptons \cite{ATLAS:2013qla} in the final state.
The final output of \texttt{CheckMATE} used to set exclusion limits in the 
MRSSM parameter space is the value for $CL_S$ of most sensitive signal region of
the studied analyses for each parameter point.

To ensure correctness of the calculated limits several tests of the used tools were done. 
For the calculation of the matrix element the event generator 
\texttt{Herwig++}-2.7  was checked against \texttt{Madgraph 5}~\cite{Alwall:2014hca},
\texttt{FeynArts}/\texttt{FormCalc} and \texttt{CalcHEP}~\cite{Belyaev:2012qa}, 
where also the input of the model files was produced by \texttt{SARAH}.
Agreement was achieved up to the implementation differences of
the programs.
The analyses implementation by \texttt{CheckMATE} was checked by re-calculating some
of the bounds given by experiment.
Additionally, the cutflow for a selection of signal regions was reproduced  
by an independent implementation within uncertainties.

With our setup we will consider the following production processes 
$pp\rightarrow \chi^0\bar{\chi}^0$, $pp\rightarrow \chi^0\chi^+,\,\chi^-\bar{\chi}^0$, 
$pp\rightarrow \chi^0\rho^-,\rho^+\bar{\chi}^0$, $pp\rightarrow \tilde{\ell}^+_R\tilde{\ell}^-_R$.
The corresponding production cross sections of our benchmark points are 
 given in Tab.~\ref{tab:EW_crosssections}. 
Processes with left-handed sleptons are neglected, assuming that the 
masses are above the detection limit as discussed before.
The masses of the relevant particles are given
in Tab.~\ref{tab:EW_mass}.

In the next sections we will show how the setup described here is used
to recast the experimental bounds on the chargino and neutralino sector of the MRSSM.
We will also highlight differences to the usual MSSM interpretations. 

\begin{table}[t!]
\centering
  \begin{tabular}{ccccccccc}
  \toprule
    process & $\chi^0\bar{\chi}^0$ & $\chi^0\chi^-$ & $\chi^0\rho^+$ & $\chi^+\bar{\chi}^0$  & $\rho^-\bar{\chi}^0$&$\chi^+\chi^-$&$\rho^+\rho^-$& $\tilde{\ell}^+_{R,i}\tilde{\ell}^-_{R,j}$\\
    \midrule
   BMP4 & 496  & 619  & 3.9  & 1147 & 0.83 & 496  & 1.25 & 17.6 \\
   BMP5 & 4.77 & 5.58 & 10.0 & 15.6 & 1.99 & 7.93 & 2.88 & $1.48\times10^{-3}$\\
   BMP6 & 1.67 & 6.04 & 22.6 & 17.0 & 5.65 & 12.1 & 8.48 & 38.4\\
   \bottomrule
  \end{tabular}

\caption{EW cross sections as calculated by \texttt{Herwig++} using the \texttt{SARAH} model input. All values given in fb. \label{tab:EW_crosssections}}
\end{table}

\begin{table}[t!]
\centering
  \begin{tabular}{cccccccccccccc}
  \toprule
   & $\chi^0_1$ & $\chi^0_2$ &  $\chi^0_3$ & $\chi^0_4$ & $\chi^\pm_1$ & $\chi^\pm_2$ & $\rho^\pm_1$ & $\rho^\pm_2$& $\tilde{\tau}_R$& $\tilde{\mu}_R$& $\tilde{e}_R$ & $\tilde{\ell}_L$ & $m_{H_1}$\\
  \midrule
  BMP4& 49.8 & 132 & 617 & 691 & 131 & 625 & 614 & 713 & 128 & 802 & 802 & 808 & 100 \\
  BMP5& 43.9 & 401 & 519 & 589 & 409 & 524 & 519 & 610 & 1000 & 1001 & 1001 & 1005 & 94\\
  BMP6& 29.7 & 427 & 562 & 579 & 422 & 562 & 433 & 587 & 106 & 353 & 353 & 508 & 95\\
   \bottomrule
  \end{tabular}

\caption{Masses of the non-SM particles in the BMPs relevant for the LHC studies discussed here. All values given in GeV.  \label{tab:EW_mass}}
\end{table}

\subsection{Light staus and light charginos}

\begin{figure}[t!]
\includegraphics[width=0.5\textwidth]{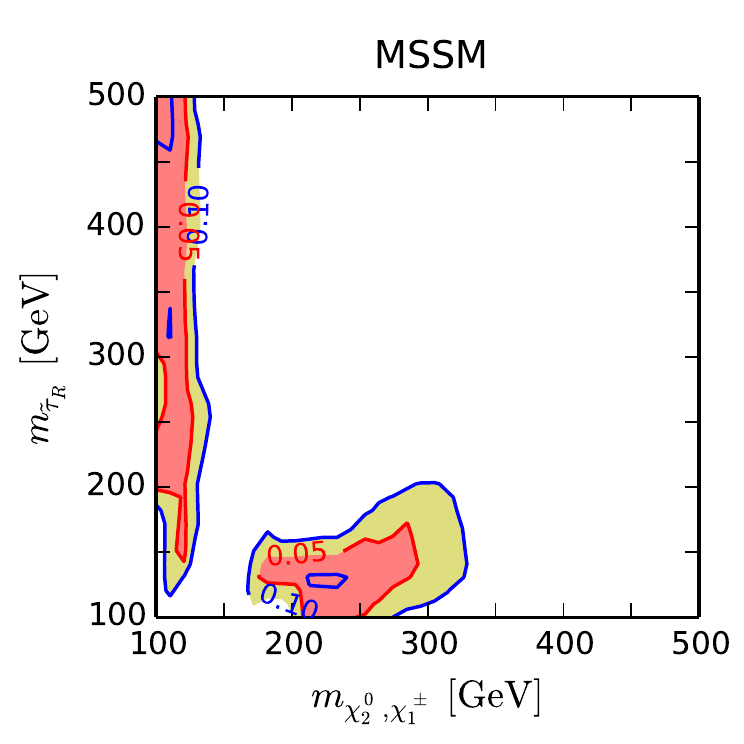}
\includegraphics[width=0.5\textwidth]{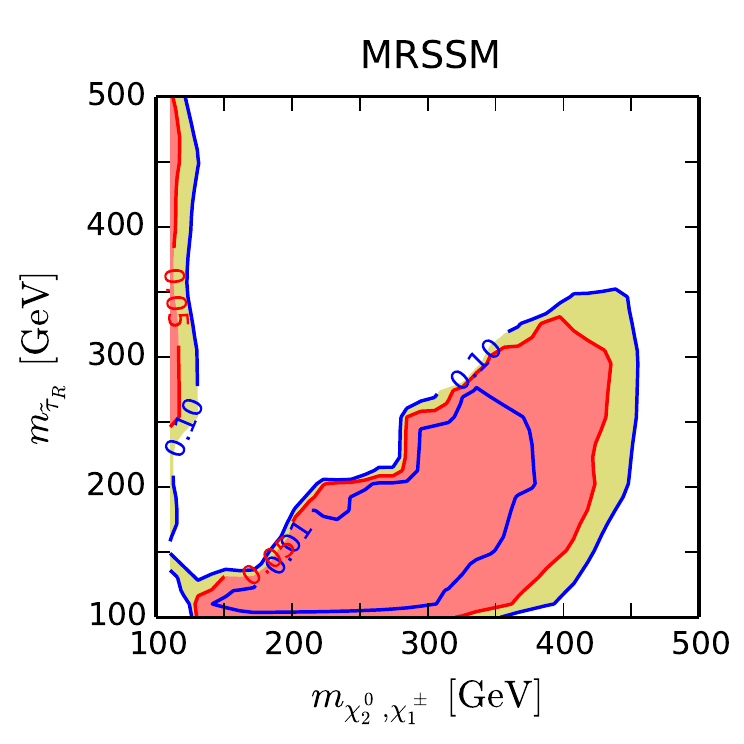}
\caption{Exclusion plots in the chargino-neutralino and stau mass plane for the first scenario of Tab.~\ref{tab:lhc-scenarios} in the MSSM (left) and in the MRSSM (right).}
\label{img:lhc_mssm_mrssm}
\end{figure}

In the following we focus on specific cases, which are of
interest for the light-singlet scenario and which are promising in
view of dark matter. We begin with the case corresponding to the first
row of Table~\ref{tab:lhc-scenarios}, i.e.~we consider a very light LSP 
(a bino-like in the MSSM, and a bino-singlino in the MRSSM), and a relatively light 
chargino (a wino in the MSSM, and a
down-(R-)higgsino in the MRSSM).  We also take relatively light staus,
which are motivated by dark matter considerations.

Fig.~\ref{img:lhc_mssm_mrssm} shows the resulting exclusion plots
taking into account the experimental analyses as discussed in
the last section.
The results show striking differences between the MSSM and the MRSSM
cases. In the MSSM the produced charginos are wino-like and all possible
decay channels (Higgs, W/Z, staus) can contribute with comparable size
to their total width.  
Therefore, as none
of the signal regions of the analyses performed by the LHC experiments have been designed 
to accept all
events/scenarios, the derived limits are rather weak. Especially for
$m_{\chi_2^0}>m_{\chi_1^0}+m_h$ the decay to Higgs channel is dominant. 
The experimental analyses used here rely on final states with charged leptons and are not
sensitive to Higgs bosons in the final state, yielding a low exclusion power.%
\footnote{Dedicated analyses taking Higgs bosons in the final state into account exist
but are not implemented in the used version of 
\texttt{CheckMATE}~\cite{Aad:2015jqa,Khachatryan:2014mma}.}
Only two small regions are
excluded: the region with a chargino mass below 100--150~GeV (depending
on the stau mass), and the region around the chargino mass of 150--300~GeV
and stau mass below 150~GeV.
In contrast, for the MRSSM the produced charginos are
down-higgsino-like and the decay to stau, if available, is
preferred. This makes the searches designed for  events with 
large multiplicity of taus very efficient. Hence, a large
triangular region with chargino mass between the decay threshold 
($m_{\tilde{\tau}_R}+m_{\text{LSP}}$) 
and around 450~GeV is excluded. Interestingly, however, to the left of
the triangle the
chargino becomes too light, decays to on-shell staus are
impossible, and decays via off-shell stau are suppressed. In that region
the acceptance is lowered and with present Run-I data no exclusion is possible.

\subsection{General investigation of light charginos}

\begin{figure}[t!]
\includegraphics{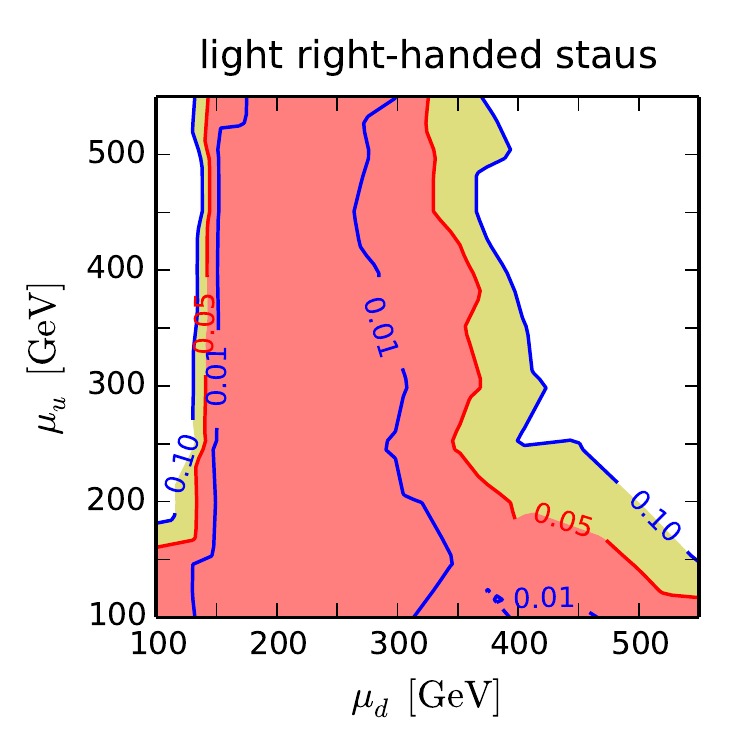}
\includegraphics{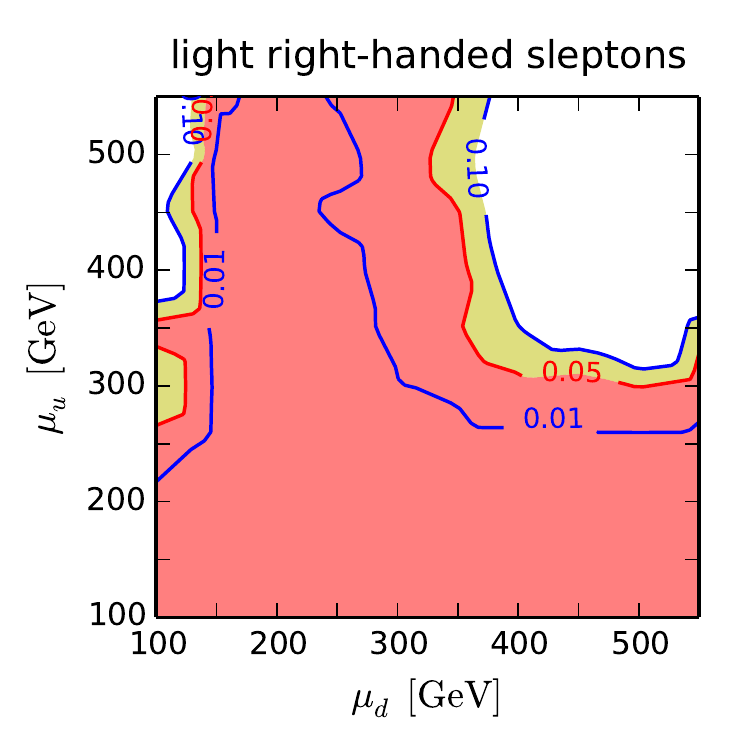}
\caption{Exclusion limits in the MRSSM for light staus as a function
of the two higgsino mass parameters $\mu_{d,u}$. 
The red (yellow) region marks the 95\% (90\%) excluded parameter
region, computed as described in the text.
The left plot has light
right-handed stau mass of 100~GeV, in the right plot all
right-handed slepton masses are set to 100~GeV. All other parameters
are fixed to the values of BMP5. Most importantly, the Dirac wino mass
is set to $M_W^D=500$~GeV.
}
\label{img:lhc_mrssm_4A}
\end{figure}

\begin{figure}[t!]
\includegraphics{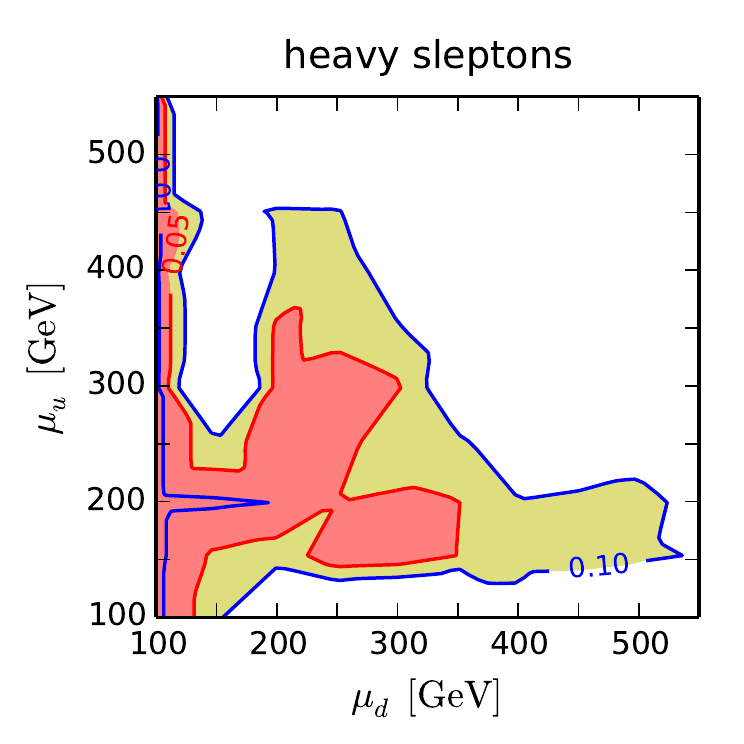}
\includegraphics{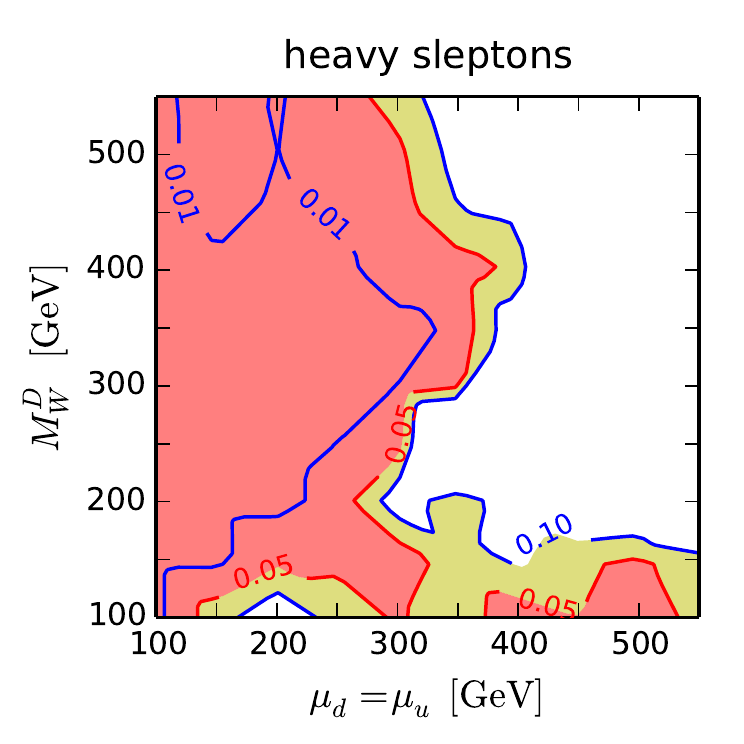}
\caption{Exclusion limits in the MRSSM for heavy sleptons as a
function of the two higgsino masses (left) and of the higgsino and
wino-triplino masses (right).
The red (yellow) region marks the 95\% (90\%) excluded parameter
region.
All parameters which are not varied in the plots are fixed to the
values of BMP5.
}
\label{img:lhc_mrssm_4B}
\end{figure}

Now we discuss the limits on chargino and neutralino masses in more
generality. Fixing the bino-singlino to be very light, $\sim50$ GeV, there are three
relevant mass parameters: the Dirac wino-triplino mass $M_W^D$ and the two
higgsino masses $\mu_{d,u}$. Further, the exclusion limits depend on
the slepton masses. 

Figure \ref{img:lhc_mrssm_4A} shows the exclusion
regions in the plane of the two higgsino masses $\mu_d$--$\mu_u$, for
the case that also right-handed staus are light. In the left plot 
all other sleptons are heavy; in the right plot the right-handed
sleptons of the other generations are also light.
In both plots there is an
interesting non-excluded region with very small $\mu_d$, below around
150~GeV. This is consistent with the discussion of the previous
subsection and with the corresponding region in
Fig.~\ref{img:lhc_mssm_mrssm}.
The region for $\mu_d$ between around 150 and 400~GeV corresponds to
the triangular region in Fig.~\ref{img:lhc_mssm_mrssm}, in which the
decay of the higgsino into stau is dominant, and is excluded.
In the case of light right-handed sleptons of all generations, the
region for the other higgsino mass $\mu_u$ below around 300~GeV is
also excluded. This originates from the mixing with the bino, leading to
non-negligible and democratically distributed decay to all
right-handed sleptons in addition to the decay to Z and Higgs boson.
Therefore selectrons and smuons will be produced in large
enough abundance to be picked up by the experiments.
As discussed before, this is in contrast to the
down-higgsino, where the mixing to the bino is suppressed via
large $\tan\beta$ and the large Yukawa coupling of tau leads to
dominating decay into staus.  
In the case of heavy sleptons of the first two generations, the
exclusion power on $\mu_u$ is very weak as the decay to Z and
Higgs boson have comparable branching ratios.
Thus the case of light right-handed staus leads to two distinct
viable parameter regions, one with very small $\mu_d$ and one with
larger $\mu_d$. These two regions are represented by the two benchmark
points BMP4 and BMP6.

Now we consider the case in which all sleptons, including the staus,
are heavy. Figure \ref{img:lhc_mrssm_4B} shows the exclusion regions
for this case, once in the $\mu_d-\mu_u$ plane (with
$M_W^D=500$~GeV), and once in the
$\mu_d$=$\mu_u - M_W^D$ plane. Both plots show that once both higgsinos are heavier 
than around 300~GeV there are parameter regions that are safe from exclusion. 
This generic, viable parameter region for heavy sleptons is
represented by benchmark point BMP5.

The left plot of Fig.~\ref{img:lhc_mrssm_4B} analyzes the case that
one of the higgsinos is lighter. It shows two additional
strips of parameter space at higgsino masses around 150~GeV, which are
not excluded by the 
considered LHC searches. The physics reason for the existence of these
regions are the higgsino decay
patterns. In the non-excluded strips the dominant decay of the light
higgsino is a two-body decay into the LSP and W/Z or SM-like Higgs
boson. Due to the small higgsino mass, the LSP is too soft to allow a
discrimination of the events from the background of standard on-shell
W/Z/Higgs production and decay. For even lower higgsino masses, the
dominant decay modes are 3-body decays via the SM bosons. In this case the
SM background for the lepton searches is smaller and the exclusion
power higher. We exemplify the region of low $\mu_d$ by the benchmark point
BMP4. For the region of low $\mu_u$ we also need to consider constraints
from dark matter searches, which will be described in the next section.

In the right of Fig.~\ref{img:lhc_mrssm_4B} 
masses of the wino-triplino $M_W^D$  below 200~GeV for 
higgsino masses above 300~GeV can not be excluded by the analyses used here.
As was discussed before, the wino-triplino decays 
predominantly into Higgs bosons and LSP and 
the Higgs boson again mainly into bottom quarks. 
As described before, this gives a low exclusion power since 
the experimental analyses so far are not sensitive to the Higgs bosons in the final state.

In summary, our investigations show that there are three
qualitatively different viable parameter regions compatible with very
light bino-singlino state, represented by the three benchmark points
BMP4,5,6. Two regions are characterized by light right-handed stau and
either light or heavy down-higgsino (where light/heavy means masses
around 100~GeV/larger than around 400~GeV, respectively). The third
region is characterized by generally heavy sleptons and higgsino
masses above around 300~GeV. In all cases, the wino-triplino mass  is not very
critical; the benchmark points have $M^D_W$ between 400--600~GeV,
but smaller values are allowed as well, and do not give rise to a different
phenomenology.

\section{Dark matter constraints}
\label{sec:dm}
As was pointed out in Sec.~\ref{sec:higgs} the scenario with a light singlet Higgs state considered
in this work leads to an upper bound on the bino-singlino mass parameter $M_B^D<55$~GeV. 
The neutralino, which is mainly given by this component, will become the LSP. Therefore, it is
the dark matter candidate of our model and we will study here how viable this scenario is when
confronted with constraints from dark matter searches.
Dirac neutralinos as a candidate for dark matter were already studied in 
Ref.~\cite{Belanger:2009wf,Buckley:2013sca}.
Especially in Ref.~\cite{Buckley:2013sca} it was clarified that the bino-singlino state is viable 
candidate as it is possible
to achieve the correct relic abundance by annihilation to leptons through sleptons,
especially a light right-handed stau.
Additionally, the spin-independent direct detection channels were investigated and
the squark and Z boson exchange as main contributions were identified. Interference between
those two channels was neglected, but we will see in the following that it plays an important role
to evade the direct detection limits.

The technical setup for the numerical computations described below is
as follows.  We use \texttt{MicrOMEGAs}-4.1.8~\cite{Belanger:2014vza} to calculate the relic density 
and direct detection rate for the MRSSM. 
The model input is given by the \texttt{CalcHEP} output of \texttt{SARAH} 
and for each parameter point the mass spectrum and couplings calculated with \texttt{SPheno}  at full one-loop level is passed to 
the program.  For the comparison to the experimental data and statistical interpretation of the 
direct detection part we use \texttt{LUXCalc}~\cite{Savage:2015xta}, which uses results from 
the LUX experiment~\cite{Akerib:2012ys}, and input on the cross section for each parameter point
from \texttt{MicrOMEGAs}.

\begin{table}
\centering
\begin{tabular}{cccc}
\toprule
   & BMP4 & BMP5 & BMP6 \\
   \midrule
$m_{\chi^1}$ &49.8 GeV&43.9 GeV&29.7 GeV\\
$m_{\tilde{\tau}_R}$ &128.5 GeV&1 TeV&105.6 GeV\\
 $\Omega h^2$  &0.119 &0.092& 0.127\\
direct detection p-value &0.9&0.5&0.2\\
\bottomrule
\end{tabular}
\caption{Values of dark matter observables for the benchmark points.}
\label{tab:bm-dm}
\end{table}

\subsection{Relic density}

We first consider the question which parameter space of the
light-singlet scenario is compatible with the measured dark matter
relic density. 
As in the MSSM, the crucial requirement is to achieve sufficiently
effective LSP pair annihilation processes. In the parameter regions
found in the previous sections, it turns out that two cases are
promising. Either, if the condition $m_{\chi_1}\approx M_Z/2$ is 
valid and S-channel resonant LSP pair annihilation into Z bosons is
possible, or if right-handed staus are light and enable annihilation
via t-channel stau exchange into tau leptons, see the corresponding Feynman diagrams in Fig.~\ref{fig:dddiagrams}, when $f,\tilde f$ replaced by $\tau,\tilde\tau$ for non-Z-exchange
diagrams.

Fig.~\ref{fig:dm} (left) shows the resulting allowed contour in the
$M_B^D$--$m_{\tilde{\tau}_R}$ parameter space. In the plot, all
non-varied parameters are fixed to the values of BMP6, but this choice
is not crucial. The important feature is that, as in the MSSM,  to meet the measured value of the relic density
the stau mass has to lie in a small interval once the
other parameters are given. The contour has a sharp resonance-like peak
around $M_B^D\approx M_Z/2$ which results from the S-channel annihilation process. 
The two different possibilities mentioned
above correspond to being at the resonance or away from it. 

In  the resonance peak the 
required stau mass has to be rather high. This is the situation realized in our
benchmark point BMP5. 
For this scenario to be viable it has to be ensured that the Z boson
does not decay to the LSP  with a detectable effect to the invisible
width of the Z boson. Since the same interaction is also 
important for, and thus constrained by, the direct detection cross
section, we will discuss it in the subsequent subsection.

For values of $M_B^D$ away from the resonance, Fig.~\ref{fig:dm}
(left) shows that the required stau
mass is small and below 150~GeV. The benchmark
points BMP4 and BMP6 represent this case. For such light staus the
LEP bound on the stau mass of $\approx 82$~GeV \cite{Agashe:2014kda}
needs to be taken into account.  In general, the 
LEP bound implies that there is a lower limit on $M_B^D$ for which the
dark matter density can be explained.

Table~\ref{tab:bm-dm} provides the precise values for the
LSP and right-handed stau  masses, as well as the
relic density, for the benchmark points. Again the different
characteristics of BMP5 and BMP4/BMP6 are visible. The benchmark point BMP4 
has been tuned to give exactly the experimental value of 
$\Omega h^2=0.1199 \pm 0.0027$~\cite{Ade:2013zuv}. For BMP5 (BMP6) the bino-singlino  (stau) mass can easily be
tuned by a small percentage to achieve full agreement with the
measured $\Omega h^2$.

\subsection{Direct detection}
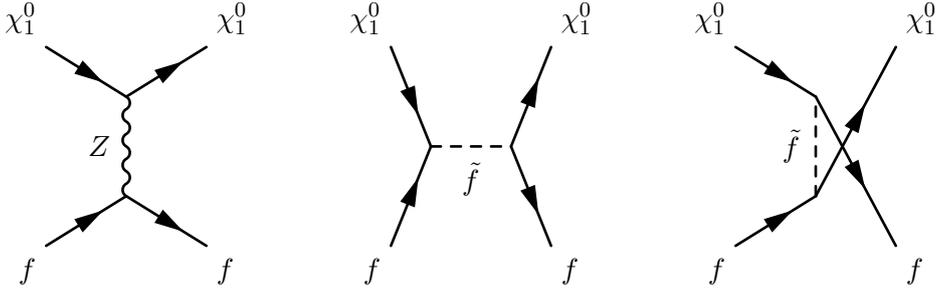
\begin{figure}[t!]
\centering
\begin{tabu} to 0.9\linewidth  {
  X[1,c,m]    
  X[1,c,m]    
  X[1,c,m]} 
\begin{fmffile}{img/Feynman/dd_Z} 
\fmfframe(20,20)(20,20){ 
\begin{fmfgraph*}(75,75) 
\fmfleft{l1,l2}
\fmfright{r1,r2}
\fmf{fermion}{l1,v1}
\fmf{fermion}{v1,r1}
\fmf{fermion}{l2,v2}
\fmf{fermion}{v2,r2}
\fmf{boson,label=$Z$}{v1,v2}
\fmflabel{$f$}{l1}
\fmflabel{$\chi^0_1$}{l2}
\fmflabel{$f$}{r1}
\fmflabel{$\chi^0_1$}{r2}
\end{fmfgraph*}} 
\end{fmffile}
&
\begin{fmffile}{img/Feynman/dd_sq} 
\fmfframe(20,20)(20,20){ 
\begin{fmfgraph*}(75,75) 
\fmfleft{l1,l2}
\fmfright{r1,r2}
\fmf{fermion}{l1,v1}
\fmf{fermion}{v2,r1}
\fmf{fermion}{l2,v1}
\fmf{fermion}{v2,r2}
\fmf{dashes,label=$\tilde{f}$}{v1,v2}
\fmflabel{$f$}{l1}
\fmflabel{$\chi^0_1$}{l2}
\fmflabel{$f$}{r1}
\fmflabel{$\chi^0_1$}{r2}
\end{fmfgraph*}} 
\end{fmffile}
&
\begin{fmffile}{img/Feynman/dd_sq2} 
\fmfframe(20,20)(20,20){ 
\begin{fmfgraph*}(75,75) 
\fmfleft{l1,l2}
\fmfright{r1,r2}
\fmf{fermion}{l1,v1}
\fmf{fermion,tension=0}{v2,r1}
\fmf{fermion}{l2,v2}
\fmf{fermion,tension=0}{v1,r2}
\fmf{phantom}{v1,r1}
\fmf{phantom}{v2,r2}
\fmf{dashes,label=$\tilde{f}$}{v1,v2}
\fmflabel{$f$}{l1}
\fmflabel{$\chi^0_1$}{l2}
\fmflabel{$f$}{r1}
\fmflabel{$\chi^0_1$}{r2}
\end{fmfgraph*}} 
\end{fmffile}
 \end{tabu}
\caption{Feynman diagrams for the most relevant LSP annihilation (with $f,\tilde f$ replaced by $\tau,\tilde\tau$) 
and direct detection processes ($f,\tilde f$ replaced by $u,\tilde u$ and $d,\tilde d$).  
\label{fig:dddiagrams}}
\end{figure}
The spin-independent dark matter--nucleon scattering cross section can be given in
terms of two scattering amplitudes $f_p,f_n$. These are conventionally
normalized such that the total spin-independent cross section at zero
momentum transfer is
\begin{equation}
  \sigma_{DM-N} = \frac{4 \mu^2_{Z_A}}{\pi}\left(Z f_p  + (A-Z)f_n\right)^2\,.
\label{eq:dm-nuc-xs}
\end{equation}
Here $\mu^2_{Z_A}$ is the dark matter--nucleon reduced mass, and $A$
and $Z$ are atomic mass and number, respectively.

As noted in Ref.~\cite{Buckley:2013sca}, the spin-independent
cross-section for Dirac neutralinos 
is dominated by the vector part of the Z boson-exchange and
squark-exchange contributions. The relevant Feynman diagrams are shown
in Fig.~\ref{fig:dddiagrams}, when $f,\tilde f$ replaced by $u,\tilde u$ and $d,\tilde d$.
Each contribution can lead to large scattering rates and thus to strong bounds on
the parameter space. For this reason we provide explicit expressions
for the relevant amplitudes.
For the Z-mediated diagram we obtain
\begin{align}
  f_p^{\text{Z boson}} & = \frac{-g_2^2}{ 8 m_Z^2 c_W^2}\left(\frac{1}{2}-2 s^2_W\right)
\left(\frac{Z_{1,3-4}+Z_{2,3-4}}{2}\right)\;,
\label{fpZ}
\\
  f_n^{\text{Z boson}} & = \frac{g_2^2}{ 16 m_Z^2 c_W^2}\left(\frac{Z_{1,3-4}+Z_{2,3-4}}{2}\right)\;,
\label{fnZ}
\end{align}
where $Z_{i,3-4} = (N^i_{13})^2-(N^i_{14})^2$ is the difference of the mixing matrix elements squared 
for the mixing of the bino with the down- and up-R-higgsino
(singlino with the down- and up-higgsino) when $i=1$ ($i=2$), see Eqs.~(\ref{eq:neut-massmatrix}) and (\ref{eq:neat-mix}).  
Here we also use the shorthand notation $s_W=\sin\theta_W$, $c_W=\cos\theta_W$.

The coupling factors $Z_{i,3-4}$ arise because the Z boson only couples to the
(R-)higgsino content of the LSP. Qualitatively, in order to suppress these amplitudes and
to ensure agreement with current direct detection bounds, the (R-)higgsino
mix-in to the LSP has
to be suppressed. This leads particularly to constraints on the higgsino mass parameters
$\mu_u$ and $\mu_d$.  As a simple illustration, we assume $\tan\beta\gg1$, $\lambda_u=0$, and $\mu_u\gg M_B^D,g_1 v$ 
and further that the direct mixing between up-higgsino and bino will
be dominant. In this case, the mixing matrix elements appearing in Eqs.~\eqref{fpZ}
and~\eqref{fnZ} take the form
\begin{equation}
Z_{1,3-4}+Z_{2,3-4}= -\left(\frac{g_1 v}{2 \mu_u}\right)^2\;,
\label{eq:neut_simple_mix}
\end{equation}
clearly exhibiting the suppression by large $\mu_u$.
Similarly, we have evaluated the squark-mediated  diagrams of
Fig.~\ref{fig:dddiagrams} for general masses of the four
first-generation squarks, $\tilde{u}_L$, 
$\tilde{u}_R$, $\tilde{d}_L$, $\tilde{d}_R$. For the qualitative 
discussion it is sufficient to provide the results in the limit of  heavy
squarks of equal mass $m_{\tilde{q}}=m_{\tilde{d}}\approx
m_{\tilde{u}}\gg m_{\chi_1}$. In this limit we obtain
\begin{align}
  f_p^{\text{heavy squark}} & = 
\frac{g_1^2}{4 m_{\tilde{q}}^2}\left[\left(\frac{4}{9}-\frac{1}{36}\right)+
\frac{1}{2}\left(\frac{1}{9}-\frac{1}{36}\right)\right]
,
\label{fphsq}
\\
  f_n^{\text{heavy squark}} & = 
\frac{g_1^2}{4 m_{\tilde{q}}^2}\left[\frac{1}{2}\left(\frac{4}{9}-\frac{1}{36}\right)
+\left(\frac{1}{9}-\frac{1}{36}\right)\right].
\label{fnhsq}
\end{align}
The minus signs result from the different Dirac structure
of the S and U channels that contribute to the process.
Therefore, in order to suppress these amplitudes and  ensure
agreement with current bounds, the squarks need to
be sufficiently heavy.%
\footnote{%
Our expressions for the direct detection amplitudes 
differ by a factor 1/16 in Eqs.~(\ref{fpZ},\ref{fnZ}) and by the relative
sign between the left-handed and right-handed squark contributions in
Eqs.~(\ref{fphsq},\ref{fnhsq}) 
from the corresponding ones of Ref.~\cite{Buckley:2013sca}.
}

For our quantitative evaluation we include the full experimental
information from LUX by using \texttt{LUXCalc} and the complete
theoretical prediction of the proton and neutron scattering amplitudes.
A likelihood is then constructed from Poisson distribution
\begin{equation}
  \mathcal{L} \left ( m_\chi, \{f_N\} | N \right ) = \frac{(\mu+b)^N e^{-(\mu+b)}}{N!} ,
\label{eq:loglike}
\end{equation}
where $N$ and $b$ are the observed and expected numbers of events by the LUX experiment, respectively, 
and $\mu$ is the expected number of signal events for a given WIMP mass and its effective couplings. 
Furthermore, we assume that DM consists in equal proportions from $\chi$ and $\bar{\chi}$ and 
use default settings for the halo profile in \texttt{MicrOMEGAs}. 

\begin{figure}
  \includegraphics{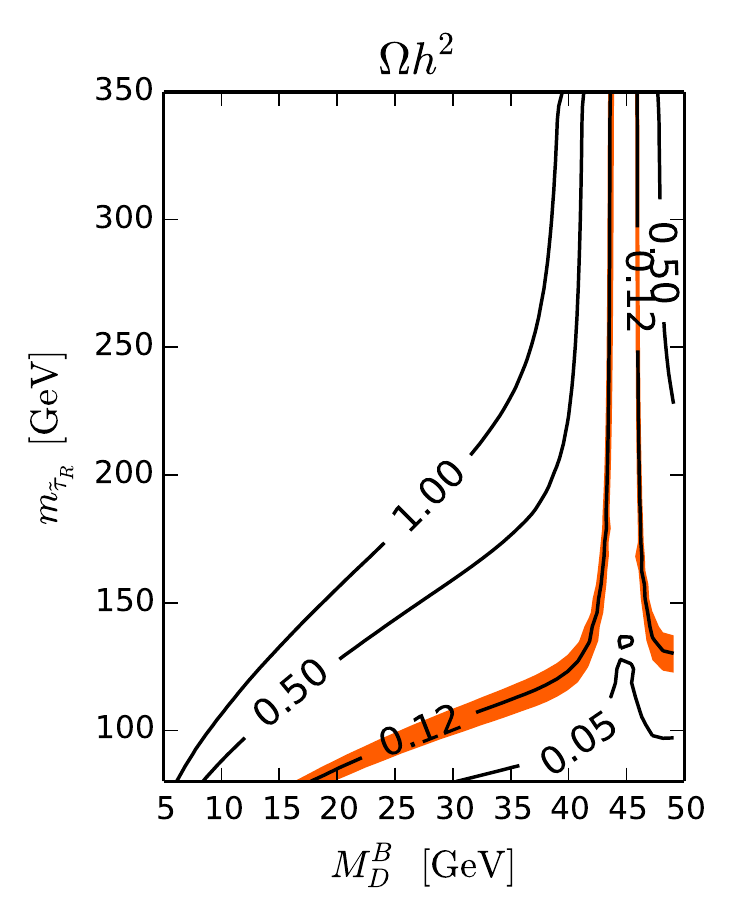}
  \includegraphics{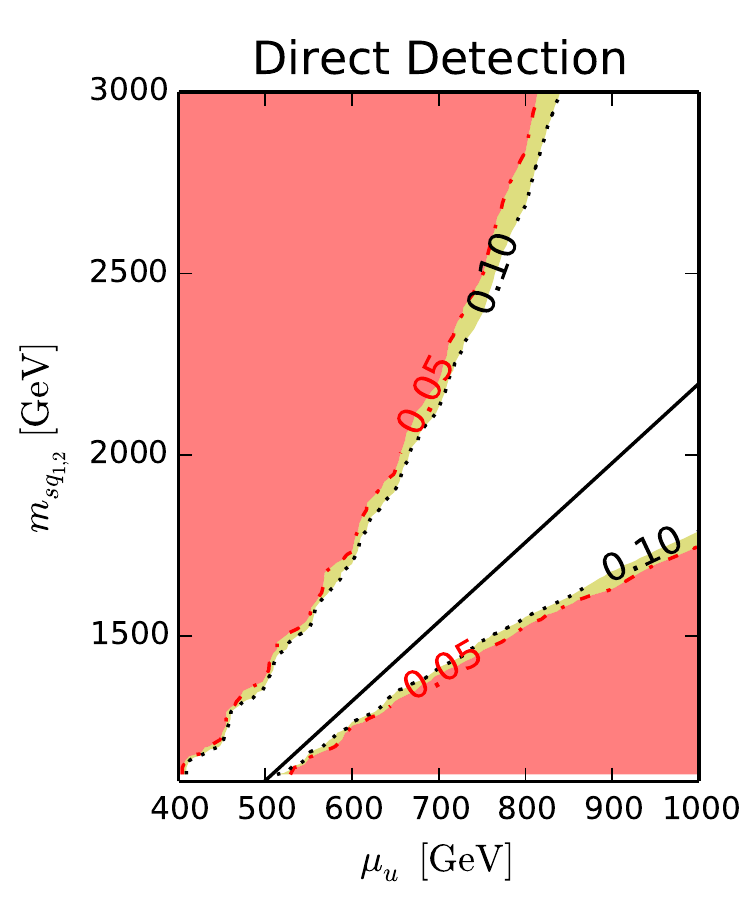}
  \caption{Dark matter relic density as a function of the bino mass parameter and right-handed slepton masses (left).
Exclusion limits in the MRSSM depending on $\mu_u$ and  equal first and second generation squark masses (right) from direct detection. 
The red with dashed dotted border (yellow with dotted border) region 
marks the 95\% (90\%) excluded parameter
region calculated  with \texttt{micrOMEGAs} and \texttt{LUXcalc}. 
The full line shows the approximation relation for the full destructive interference in Eq.~\eqref{eq:dd_not_simple_mass_rel}. 
The non-varied parameters in each plot are fixed to
the values of BMP6.}
\label{fig:dm}
\end{figure}

On the right of Fig.~\ref{fig:dm}, the 95~\% and 90~\% exclusion bounds (violet (dark) and yellow (light dark) regions) 
derived using the log likelihood for the direct detection
by LUX given in Eq.~\eqref{eq:loglike} are shown
depending on $\mu_u$ and the first/second generation squark masses.
Here, it can be seen that the derived limits are quite sensitive to the
combination of both parameters. 
This dependence stems from an interference of the amplitudes
and can be understood by a simplified analysis of the expressions, as given before.

In the approximation leading to Eq.~\eqref{eq:neut_simple_mix}, we can add the Z- and 
squark-exchange contributions to a simple form
 \begin{equation}
f_n^{\text{Z boson}+\text{heavy squark}} = \frac{g_1^2}{96}\left(\frac{7}{ m_{\tilde{q}}^2}-\frac{3}{\mu_u^2}\right),\quad
f_p^{\text{Z boson}+\text{heavy squark}} = \frac{g_1^2}{96}\left(\frac{11}{ m_{\tilde{q}}^2}\right),
\end{equation}
where $s_W^2=1/4$ has been taken in the proton case.
Plugging these results into the expression for the dark matter--nucleus cross section,
Eq.~\eqref{eq:dm-nuc-xs},
we find a complete destructive interference, i.e. $\sigma_{DM-N}=0$, when
  \begin{equation}
m_{\tilde{q}}= \sqrt{\frac{7+11 \frac{A}{Z-A}}{3}}\mu_u \overset{\text{Xe}}{\approx} 2.2\mu_u\;,
\label{eq:dd_not_simple_mass_rel}
\end{equation}
where the numerical value for xenon is calculated with  $A=54$ and $Z=131.3$. 

The line of destructive interference corresponding to
Eq.~\eqref{eq:dd_not_simple_mass_rel}, shown as a full line in
Fig.~\ref{fig:dm} (right), 
explains the funnel-shaped allowed region.
Above the line, the squark-mediated amplitudes become small, to the right of the line, the Z-mediated
amplitudes become small.
It should be noted that the result for the exclusion bounds is calculated using the complete 
information of \texttt{micrOMEGAs} and \texttt{LUXCalc},
while Eq.~\eqref{eq:dd_not_simple_mass_rel} is only approximated.
As the squark masses of the first two generations are limited by LHC searches, the direct detection
non-discovery provides a lower limit on $\mu_u$. 
This also excludes the region of low $\mu_u$ allowed by LHC searches 
in the left of Fig.~\ref{img:lhc_mrssm_4B}.

As discussed before, the amplitude for the Z-mediated exchange is closely related to one responsible for the
Z boson decay to a pair of LSP. Both depend strongly on the $Z_{i,3-4}$ containing the neutralino mixing matrix elements.
The direct detection limit of Fig.~\ref{fig:dm} limits the $Z_{i,3-4}$ strongly enough so that 
$\Gamma(Z\rightarrow\chi^0_{1}\bar{\chi}^0_{1})\ll 3$~MeV is ensured over all the parameter space. 

Xenon is not the only element used for direct detection experiments, therefore
the change of the values of $A$ and $Z$ has to be taken into account, when using
Eq.~\eqref{eq:dd_not_simple_mass_rel} for other cases.
The variation of the ratio $A/(Z-A)$ ($0.7$ for xenon, $0.8$ for argon and germanium)
is small enough that the regions of strong destructive interference for different
experiments will not overlap in the MRSSM parameter space.
Here, we only take the LUX result into account as it is the most sensitive one available.

\section{Summary and conclusions}
\label{sec:results}
\begin{figure}[th!]
(a)\includegraphics{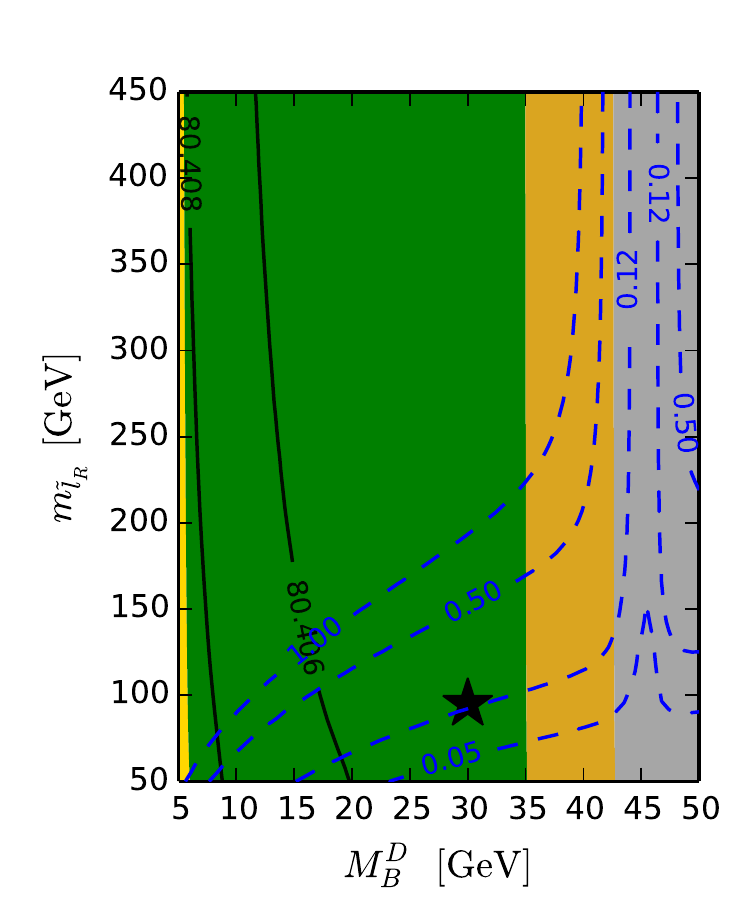}
(b)\includegraphics{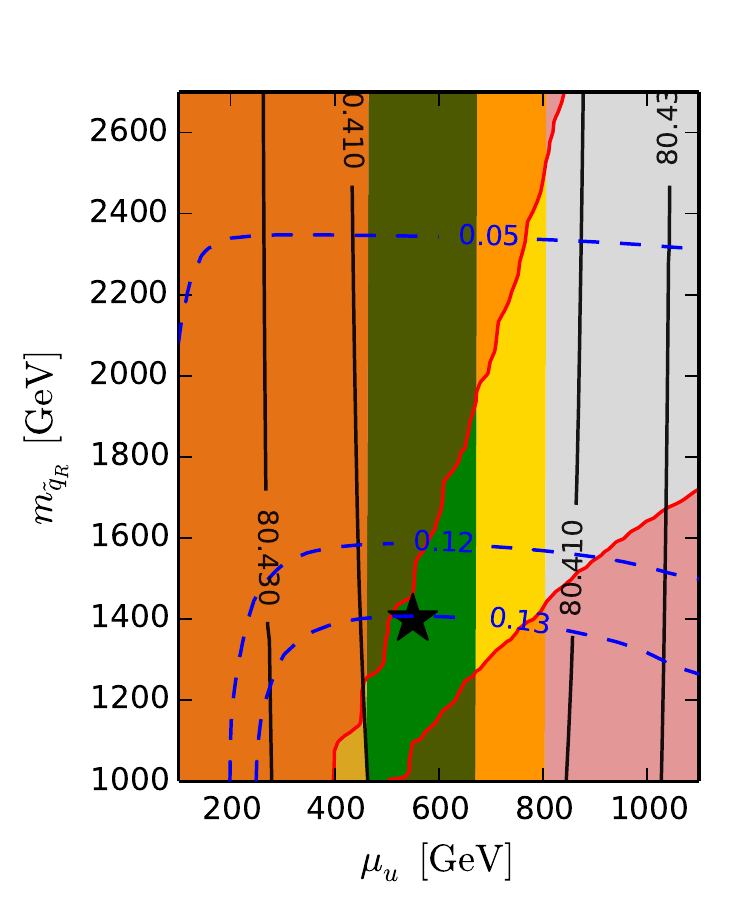}\\
(c)\includegraphics{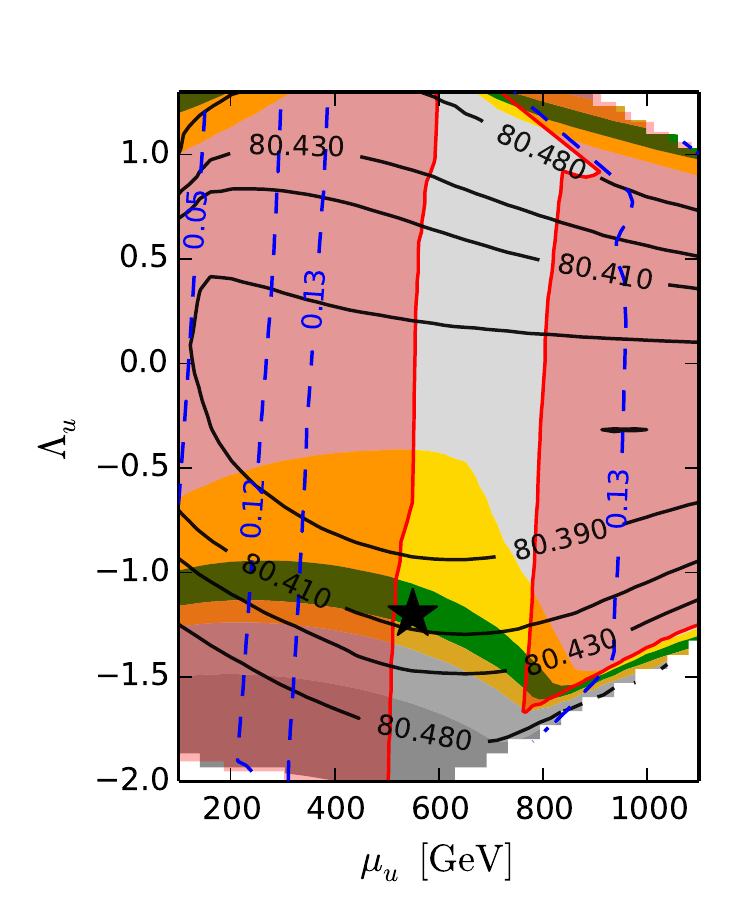}
(d)\includegraphics{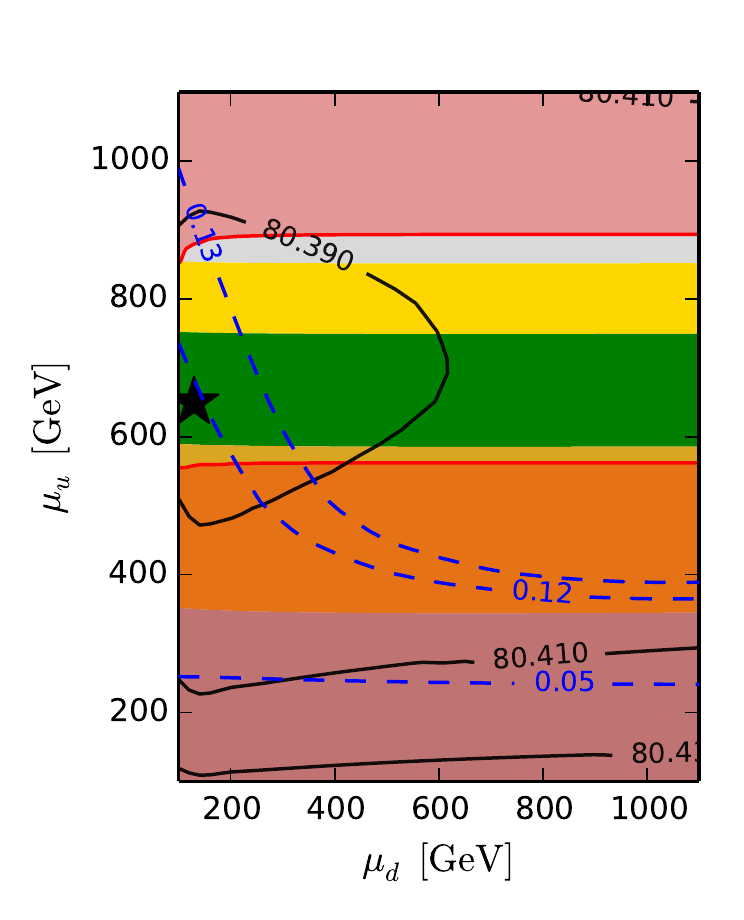}\\
 \includegraphics{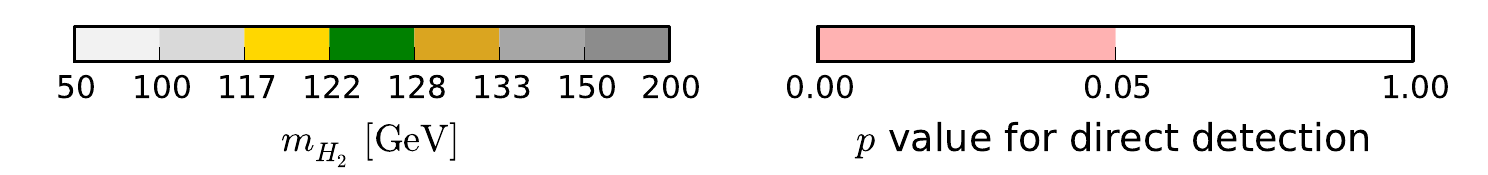}
\caption{Shown are parameter regions and their agreement with experiment.
Red overlaid areas are  excluded with 95\% CL by dark matter direct detection.
Black full (blue dashed) lines show  $m_W$ ($\Omega h^2$). The mass of the
SM-like Higgs boson $m_{H_2}$ is given by the colour scale shown. All non-varied parameters are
set to the values of BMP6, except for the bottom right plot, where BMP4 is used. BMPs are marked by stars.
The white areas in (d) stem from tachyonic states appearing in the mass spectrum,
at the borders to these areas the interpolation breaks down.}
\label{img:2dplots1}
\end{figure}
The minimal R-symmetric model MRSSM is a promising alternative to the
MSSM, which predicts Dirac gauginos and higgsinos as well as scalars
in the adjoint representation of
SU(3)$\times$SU(2)$\times$U(1). Here we have investigated the
possibility that the scalar singlet mass is small and gives rise to a
singlet-like mass eigenstate with mass below 125~GeV. The
potential advantages of this scenario are an increased tree-level
SM-like Higgs boson mass, many light weakly interacting particles
which could be discovered at the next LHC run, and the possibility to
explain dark matter.

Many of the model parameters are strongly constrained in order to make
this scenario viable. Here we briefly summarize the main
constraints. Fig.~\ref{img:2dplots1} summarize
the four most important non-LHC constraints on the relevant parameters.
There, the SM-like Higgs boson mass is given by the green-yellow colour bands, while
the red area shows the regions excluded by direct detection searches. Black full (blue dashed) 
lines give the W boson mass ($\Omega h^2$) contours.
All plots are based on BMP6, except for Fig.~\ref{img:2dplots1}(d), where BMP4 was used.

In order to realize a light singlet-like Higgs boson in the first
place, not only the parameter $m_S$ but also $M_B^D$ and $\lambda_u$ must
be very small. The SM-like Higgs boson
mass is increased at tree-level compared to the case with heavy
singlet or to the MSSM, but loop contributions governed by
$\Lambda_{u}$ are still important. 
Accordingly, the plots of Fig.~\ref{img:2dplots1}
show that the Higgs mass measurement essentially fixes $\Lambda_u$ and
constrains $M_B^D$ and $\mu_u$. We have found that the other constraints in Higgs
physics, arising from measurements of Higgs properties and of
non-discovery of further Higgs states, can be easily fulfilled.
As discussed in
Ref.~\cite{Diessner:2014ksa}, the W-boson mass can be explained
simultaneously with the SM-like Higgs boson
mass. Fig.~\ref{img:2dplots1} confirms that this is still the case in
the light singlet scenario.

The dark matter relic density can be explained in the MRSSM light
singlet scenario because the Dirac bino-singlino neutralino is necessarily
the stable LSP. Agreement with the observed relic density requires 
particular combinations of the bino-singlino mass $M_B^D$ and the
right-handed stau mass. If the latter is heavy, as in our benchmark
point BMP5, the LSP 
mass must be close to half of the Z boson mass. If the LSP mass is
away from the Z boson resonance, the right-handed
stau must be light, as in our BMP4 and BMP6. This resonance behaviour is
clearly displayed in 
Fig.~\ref{img:2dplots1}(a). Parameters like $\mu_u$, $\mu_d$ influence
the relic density due to their impact on the LSP mass via mixing, Fig.~\ref{img:2dplots1}(d).

Further bounds on the model parameters arise from the negative
searches for dark matter and from the negative LHC searches for SUSY
particles.
The direct dark matter searches correlate $\mu_u$ with $m_{\tilde{q}_R;1,2}^2$, Fig.~\ref{img:2dplots1}(b). 
Because of the non-discovery of squarks at LHC so far, this give rise to a lower bound on $\mu_u$, 
which is not affected by other parameters of the weak sector like $\Lambda_u$, Fig.~\ref{img:2dplots1}(c).  
Together with the Higgs mass value, 
however, these searches constrain both $\mu_u$ and $m_{\tilde{q}_R;1,2}^2$ to a rather 
narrow range, see Fig.~\ref{img:2dplots1}(b). 

Our recast of LHC analyses has revealed three distinct
viable parameter regions of interest, represented by the three
benchmark parameter points BMP4, BMP5, BMP6. The most obvious region is
characterized by heavy $M_W^D$, $\mu_d$ and sleptons (BMP5). 
A second is characterized by very light right-handed
stau mass around 100~GeV (BMP6); in the third region the right-handed
stau and $\mu_d$ are very light (BMP4). Fig.~\ref{img:2dplots1}(d), which
gives the parameter variation for BMP4, shows that this region is also allowed by the other constraints. 
As explained in Sec.~\ref{sect:LHC}, these two regions are allowed because the
existing LHC searches become ineffective. The LHC searches alone would
allow further parameter regions with very small $\mu_u$, which however
are excluded by dark matter constraints.

To summarize: the experimental data from collider and dark matter experiments impose  
stringent constraints on the parameter of the light scalar scenario of the MRSSM. 
Nevertheless we have identified regions in the parameter space and proposed representative 
benchmark points fulfilling all the constraints. All viable parameter
regions are characterized by several light weakly interacting SUSY
particles and will be tested both by future dark matter and LHC SUSY searches.

\acknowledgments
Work supported in part by the German DFG Research Training Group 1504 and the DFG grant STO 876/4-1, 
the Polish National Science Centre grants under OPUS-2012/05/B/ST2/03306 and 
the European Commission through the contract PITN-GA-2012-316704 (HIGGSTOOLS).
\newline

\noindent WK would like to thank Aleksandra Drozd for useful discussions concerning dark matter.
PD would like to thank Daniel Schmeier for technical support with \texttt{CheckMATE}.
We thank Christopher Savage for help on \texttt{LUXCalc} usage in the case of Dirac dark 
matter candidate.


\appendix
\section{MRSSM mass matrices at tree level}
\label{sect:app}
For completeness we recall here the tree-level mass matrices for the Higgs bosons and the charginos and neutralinos of the MRSSM, which are relevant for the discussion of the present paper.\\

\noindent a) Pseudo-scalar Higgs bosons:\\
Since in the pseudo-scalar sector of $(\sigma_d,\sigma_u,\sigma_S,\sigma_T)$ there is no mixing between the MSSM-like states $(\sigma_d,\sigma_u)$ and the singlet-triplet states $(\sigma_S,\sigma_T)$, the mass-squared matrix breaks into two 2x2 submatrices as follows
\begin{equation}
\mathcal{M}^\sigma_{u,d}=
\begin{pmatrix}
B_\mu\frac{v_u}{v_d} & B_\mu \\
B_\mu & B_\mu\frac{v_d}{v_u}\\
\end{pmatrix},
\qquad
\mathcal{M}^\sigma_{S,T}=
\begin{pmatrix}
 m_S^2+\frac{\lambda_d^2 v_d^2+\lambda_u^2 v_u^2 }{2} & \frac{\lambda_d\Lambda_d v_d^2-\lambda_u\Lambda_u v_u^2}{2 \sqrt{2}} \\
\frac{\lambda_d\Lambda_d v_d^2-\lambda_u\Lambda_u v_u^2}{2 \sqrt{2}}& m_T^2+ \frac{\Lambda_d^2 v_d^2+\Lambda_u^2 v_u^2 }{4}\\
\end{pmatrix}.
\end{equation}

\noindent b) Neutralinos:\\
In the weak basis of eight neutral  electroweak two-component fermions: 
$ {\xi}_i=({\tilde{B}}, \tilde{W}^0, \tilde{R}_d^0, \tilde{R}_u^0)$,  
$\zeta_i=(\tilde{S}, \tilde{T}^0, \tilde{H}_d^0, \tilde{H}_u^0) $ 
with R-charges $+1$, $-1$ respectively, the neutralino mass matrix takes the form of
\begin{equation} 
\label{eq:neut-massmatrix}
m_{{\chi}} = \left( 
\begin{array}{cccc}
M^{D}_B &0 &-\frac{1}{2} g_1 v_d  &\frac{1}{2} g_1 v_u \\ 
0 &M^{D}_W &\frac{1}{2} g_2 v_d  &-\frac{1}{2} g_2 v_u \\ 
- \frac{1}{\sqrt{2}} \lambda_d v_d  &-\frac{1}{2} \Lambda_d v_d  & - \mud{+} &0\\ 
\frac{1}{\sqrt{2}} \lambda_u v_u  &-\frac{1}{2} \Lambda_u v_u  &0 & \muu{-}
\end{array} 
\right) .
 \end{equation} 
The transformation to a diagonal mass matrix and mass eigenstates $\kappa_i$ and $\psi_i$ 
is performed by two unitary mixing matrices \(N^1\) and \(N^2\) as
\begin{align} \nonumber
N^{1,*} m_{{\chi}} N^{2,\dagger} &= m^{diag}_{{\chi}} \,,
&
{\xi}_i&=\sum_{j=1}^4 N^{1,*}_{ji}{\kappa}_j\,,
&
\zeta_i=\sum_{j=1}^4 N^{2,*}_{ij}{\psi}_j\,,
\end{align} 
and physical four-component Dirac neutralinos are constructed as
\begin{equation}
{\chi}^0_i=\left(\begin{array}{c}
\kappa_i\\
{\psi}^{*}_i\end{array}\right)\qquad i=1,2,3,4.
\label{eq:neat-mix}
\end{equation}

\noindent c) Charginos:\\
The mass matrix of charginos in the weak basis of eight charged two-component
fermions breaks  into two 2x2 submatrices. 
The first, in the
basis \( (\tilde{T}^-, \tilde{H}_d^-), (\tilde{W}^+, \tilde{R}_d^+) \)
of spinors with R-charge equal to electric charge,
 takes the form of
\begin{equation} 
m_{{\chi}^+} = \left( 
\begin{array}{cc}
g_2 v_T  + M^{D}_W \; &\frac{1}{\sqrt{2}} \Lambda_d v_d \\ 
\frac{1}{\sqrt{2}} g_2 v_d \; &+ \mud{-}
\end{array} 
\right) 
\label{eq:cha1-massmatrix}
 \end{equation} 
The diagonalization and transformation to mass eigenstates $\lambda^\pm_i$ is
performed by two unitary matrices \(U^1\) and \(V^1\)  as
\begin{align} 
U^{1,*} m_{{\chi}^+} V^{1,\dagger} &= m^{diag}_{{\chi}^+} \,,
&\tilde{T}^- &= \sum_{j=1}^2U^{1,*}_{j 1}\lambda^-_{{j}}\,,& 
\tilde{H}_d^- &= \sum_{j=1}^2U^{1,*}_{j 2}\lambda^-_{{j}}\,,\\ 
&&
\tilde{W}^+ &= \sum_{j=1}^2V^{1,*}_{1 j}\lambda^+_{{j}}\,,&
R_d^+ &= \sum_{j=1}^2V^{1,*}_{2 j}\lambda^+_{{j}}
\end{align} 
and the corresponding physical four-component charginos are built as 
\begin{equation}
{\chi}^+_i=\left(\begin{array}{c}
\lambda^+_i\\
\lambda^{-*}_i\end{array}\right)\qquad i=1,2.
\end{equation}
The second submatrix, in the basis $ (\tilde{W}^-, R_u^-)$, $(\tilde{T}^+, \tilde{H}_u^+) $
 of spinors with R-charge equal to minus electric charge, reads
\begin{equation} 
m_{{\rho}^-} = \left( 
\begin{array}{cc}
- g_2 v_T  + M^{D}_W \;&\frac{1}{\sqrt{2}} g_2 v_u \\ 
- \frac{1}{\sqrt{2}} \Lambda_u v_u \; &  - \muu{+} \end{array} 
\right) 
\label{eq:cha2-massmatrix}
 \end{equation} 
The diagonalization and transformation to mass eigenstates $\eta^\pm_i$  is
performed by \(U^2\) and \(V^2\) as 
\begin{align} 
U^{2,*} m_{{\rho}^-} V^{2,\dagger} & = m^{diag}_{{\rho}^-} \,,&
\tilde{W}^- & = \sum_{j=1}^2U^{2,*}_{j 1}\eta^-_{{j}}\,, &
R_u^- & = \sum_{j=1}^2U^{2,*}_{j 2}\eta^-_{{j}}\\ 
&&\tilde{T}^+ & = \sum_{j=1}^2V^{2,*}_{1 j}\eta^+_{{j}}\,, &
\tilde{H}_u^+ & = \sum_{j=1}^2V^{2,*}_{2 j}\eta^+_{{j}}
\end{align} 
and the other two physical four-component charginos are built as 
\begin{equation}
{\rho}^-_i=\left(\begin{array}{c}
\eta^-_i\\
\eta^{+*}_i\end{array}\right)\qquad i=1,2.
\end{equation}

\end{document}